\documentclass[preprint]{IEEEtran}
\usepackage{supertabular} % tableaux qui tiennent sur plusieurs pages
\usepackage{graphicx}
\usepackage{lineno,hyperref}
\usepackage{amsmath}
\usepackage{amsfonts}
\usepackage{amssymb}
\usepackage{array,multirow,makecell}
\modulolinenumbers[5]
\usepackage{lscape}
\begin{document}
\title{Privacy-preserving schemes for Ad Hoc Social Networks: A survey}
\author{Mohamed~Amine~Ferrag,
        Leandros~Maglaras,~\IEEEmembership{Senior Member,~IEEE,}
        and~Ahmed~Ahmim % <-this % stops a space
\thanks{M. A. Ferrag is with Department of Computer Science, Guelma University, Algeria e-mail: mohamed.amine.ferrag@gmail.com  phone: +213661-873-051}% <-this % stops a space
\thanks{L. Maglaras is with School of Computer Science and Informatics, Cyber Security Centre, De Montfort University, Leicester, UK e-mail: leandros.maglaras@dmu.ac.uk}% <-this % stops a space
\thanks{A. Ahmim is with Department of Mathematics and Computer Science, University of Larbi Tebessi, Algeria e-mail: a.ahmim@gmail.com}% <-this % stops a space
\thanks{Manuscript received 2016}}
\onecolumn


\markboth{IEEE 
}%
{Shell \MakeLowercase{\textit{et al.}}: Bare Demo of IEEEtran.cls for IEEE Communications Society Journals}



\maketitle

\begin{abstract}
In this paper, we review the state of the art of privacy-preserving schemes for ad hoc social networks, including, mobile social networks (MSNs) and vehicular social networks (VSNs). Specifically, we select and in-detail examine thirty-three privacy-preserving schemes developed for or applied in the context of ad hoc social networks. These schemes are published between 2008 and 2016. Based on this existing privacy preservation schemes, we survey privacy preservation models, including location privacy, identity privacy, anonymity, traceability, interest privacy, backward privacy, and content oriented privacy. The recent important attacks of leaking privacy, countermeasures, and game theoretic approaches in VSNs and MSNs are summarized in form of tables. In addition, an overview of recommendations for further research is also provided. With this survey, readers can have a more thorough understanding of research trends in privacy-preserving schemes for ad hoc social networks.
\end{abstract}


% \fi
\IEEEpeerreviewmaketitle

\section{Introduction}
\IEEEPARstart{A}{wireless ad-hoc} network consists of mobile platforms which are free to communicate without any central control entity \cite{haas2002wireless}. It can operate in an isolated manner or with fixed networks through gateways. The power of an ad hoc network is that it does not differentiate  between a router and a station, i.e., each station contributes to routing. Mobile ad hoc networks (MANETs) and vehicular ad hoc networks (VANETs) are special cases of ad hoc networks. The MANET is an autonomous system of mobile nodes, which has several salient characteristics, namely, dynamic topologies, bandwidth-constrained and energy constrained operation, and limited physical security \cite{100, 101}. The VANET is a special case of MANET, where the mobile nodes are instantiated with vehicles equipped with On-board Unit (OBU) communication devices \cite{57}.

Today, in our daily lives,  social networking enables us to contact our colleagues, friends, and families through applications such as Facebook, Twitter, Linkedin, Google+, YouTube, and ResearchGate. At the same time, however,  ad hoc social networks are getting increasingly important which it takes the \textit{human factors} into consideration, i.e., human mobility, human selfish status, and human preferences \cite{14,15,16,17}. In this survey, we focus on two types of ad hoc social networks, including, mobile social networks (MSNs) and vehicular social networks (VSNs). Tab. \ref{tab:Tab1} gives a comparison between ad hoc network (MANET, VANET) and ad hoc social network (MSN, VSN) in terms of node, mobility, connectivity, resource, architecture, scalability, application, typical research issue, and security.

\begin{table}[h]
\centering
\caption{Comparison of network characteristics}
\begin{tabular}{|p{0.8in}||p{0.9in}|p{0.9in}||p{1in}|p{1in}|} \hline 
\textbf{} & \multicolumn{2}{|p{1.3in}||}{\textbf{Ad hoc network}} & \multicolumn{2}{|p{1.7in}|}{\textbf{Ad hoc social network}} \\ \hline 
\textbf{} & \textbf{MANET} & \textbf{VANET} & \textbf{MSN} & \textbf{VSN} \\ \hline 
\textbf{Node} & Laptop, smartphone, pocket PC, router\dots etc & Vehicles & Human & Vehicles with social properties \\ \hline 
\textbf{Mobility} & Random & On-Road & Human mobility & On-Road with socialspot \\ \hline 
\textbf{Connectivity} & Random & Random and Intermittent & Interest social & Interest social \\ \hline 
\textbf{Resource} & Limited hardware and power limited by battery & Almost unlimited & Limited hardware and power limited by battery & Almost unlimited \\ \hline 
\textbf{Architecture} & Node-to-node & Vehicle-to-vehicle, Vehicle-to-RSU & Node-to-node , Node-to-socialspot & Vehicle-to-vehicle, Vehicle-to-RSU, Vehicle-to-socialspot \\ \hline 
\textbf{Scalability} & 50-100 nodes & Huge & 50-100 nodes & Huge \\ \hline 
\textbf{Application} & Military, disaster (specific) & Safety, traffic, payment & Mobile social applications, location-based applications & Vehicular social applications, location-based applications \\ \hline 
\textbf{Typical Research Issue} & Routing & Application & Application & Application \\ \hline 
\textbf{Security} & Sensitive & Sensitive & Highly-sensitive & Highly-sensitive \\ \hline 
\end{tabular}
\label{tab:Tab1}
\end{table}
\begin{table}[h]
\centering
\caption{Publication date breakdown - Surveyed papers}
\begin{tabular}{|p{0.5in}|p{0.5in}||p{0.5in}|p{0.5in}||p{0.5in}|p{0.5in}|} \hline 
\textbf{Paper\newline } & \textbf{Year} & \textbf{Paper} & \textbf{Year} & \textbf{Paper} & \textbf{Year} \\ \hline 
\cite{34} & 2008 & \cite{27} & 2011 & \cite{24} & 2013 \\ \hline 
\cite{20} & 2010 & \cite{29} & 2011 & \cite{5} & 2014 \\ \hline 
\cite{28} & 2010 & \cite{30} & 2011 & \cite{6} & 2014 \\ \hline 
\cite{31} & 2010 & \cite{8} & 2012 & \cite{21} & 2014 \\ \hline 
\cite{32} & 2010 & \cite{9} & 2012 & \cite{13} & 2015 \\ \hline 
\cite{33} & 2010 & \cite{18} & 2012 & \cite{118} & 2015 \\ \hline 
\cite{10} & 2011 & \cite{19} & 2012 & \cite{51} & 2016 \\ \hline 
\cite{11} & 2011 & \cite{22} & 2012 & \cite{50} & 2016 \\ \hline 
\cite{12} & 2011 & \cite{23} & 2012 & \cite{7} & 2016 \\ \hline 
\cite{25} & 2011 & \cite{1} & 2013 & \cite{3} & 2016 \\ \hline 
\cite{26} & 2011 & \cite{4} & 2013 & \cite{102} & 2016 \\ \hline 
\end{tabular}
\label{tab:Tab2}
\end{table}
\begin{figure}[h]
\centering
\includegraphics[width=0.8\linewidth]{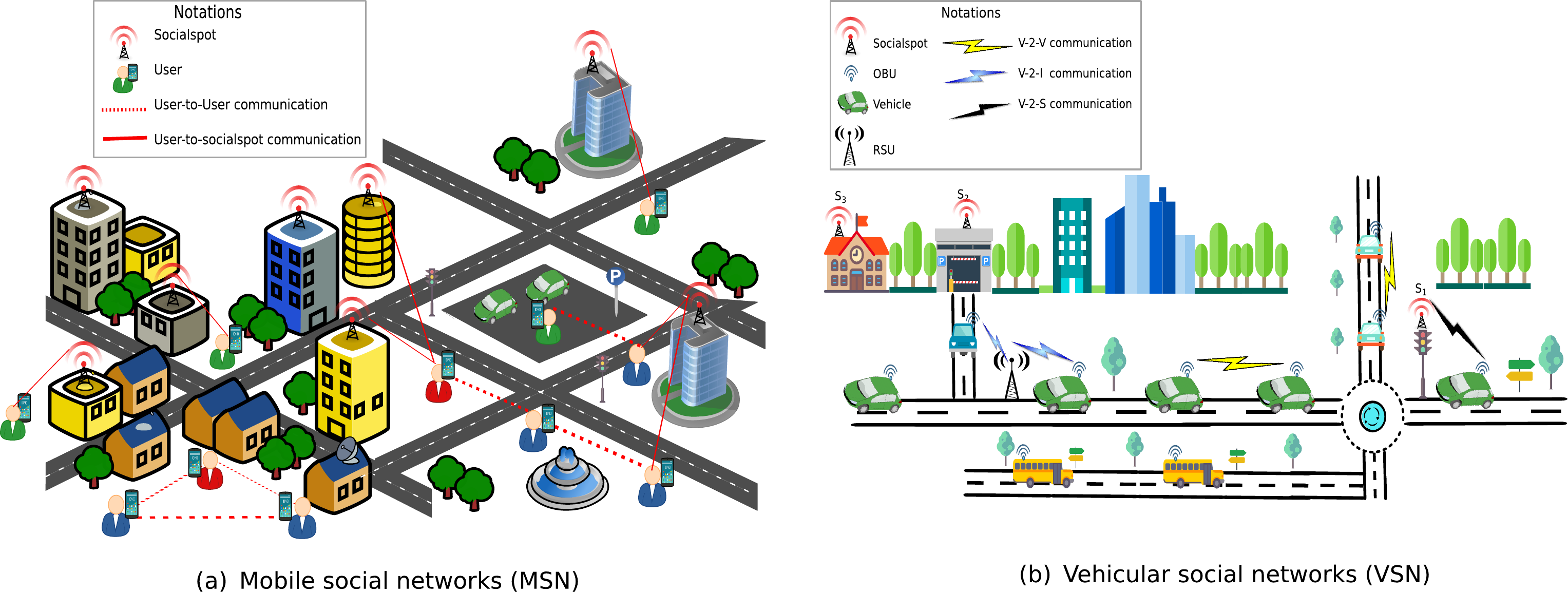}
\caption{A global architecture of Ad Hoc Social Network}
\label{fig:Fig1}
\end{figure}
\begin{figure}[h]
\centering
\includegraphics[width=0.8\linewidth]{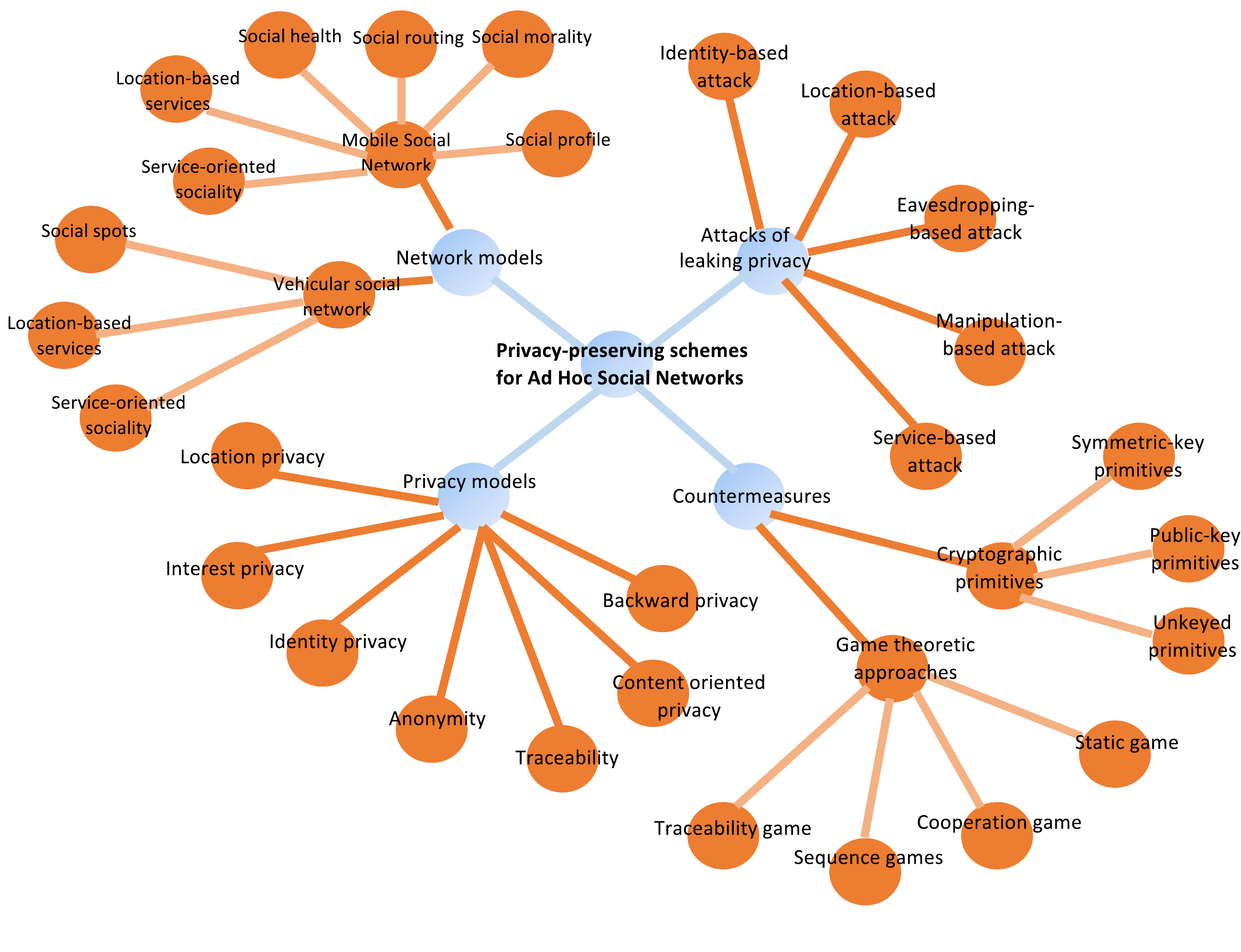}
\caption{Organization of surveyed papers}
\label{fig:Fig2}
\end{figure}

As shown in Fig.\ref{fig:Fig1}.a, MSN is composed of a set of mobile users $U=\{u_1,\cdots ,u_n\}$ with some socialspots $S=\{s_1,\cdots ,s_n\}$ in a city environment, which each user $u_i$ having equal communication range $R_{u_i}$ \cite{14,16,17}. Similarly to MSN, as shown in Fig.\ref{fig:Fig1}.b, a VSN is composed of a large number of vehicles $V=\{v_1,\cdots ,v_n\}$ equipped with on-board units (OBUs), Roadside Units (RSUs) and some socialspots $S=\{s_1,\cdots ,s_n\}$ . Using the communication capabilities of their OBUs, the vehicles can communicate with each other, as well as with RSUs and socialspot $s_i$, i.e., vehicle-to-vehicle (V-2-V) communication, vehicle-to-infrastructure (V-2-I) communication, vehicle-to-socialspot (V-2-S) communication \cite{15,76}. These three types of communications can happen using the wireless access technologies that are available today such as cellular systems (3G/4G/5G), WLAN/Wi-Fi, WiMAX, and DSRC/WAVE \cite{119}.

Since the social characteristics are integrated into ad hoc networks, MAN and VSN have become very sensitive to security and privacy compared with MANET and VANET. In other words, security issue is crucial to the full adoption of MSN and VSN, especially for the privacy. Based on identities, pseudonyms, locations, and profiles, an adversary can launch active or passive attacks (deliberately delays, drops, corrupts, or modifies messages) in order to steal the social data as well as to damage the V-2-V, V-2-I, and V-2-S communications. Hence, in order to protect the social community, the privacy-preserving schemes employed in MSNs and VSNs should satisfy the following security requirements: authentication, integrity, non-repudiation, access control, and privacy \cite{106}.

In a recent survey paper which was published in 2016 \cite{116} authors  review the state of the art of privacy preserving techniques for online social networks (OSNs). In addition, other recent surveys in \cite{75} and \cite{76} published in 2015 provide information about applications, platforms, system architectures in MSNs and VSNs, respectively. Neither of the published, state-of-the-art literature provide a comprehensive survey for recent advances in privacy-preserving schemes for MSNs and VSNs. The aim of this survey paper is to provide comprehensive and systematic review of the recent studies on published privacy-preserving schemes for ad hoc social networks, including, MSNs and VSNs. More precisely, we select and in-detail examine thirty-three privacy-preserving schemes. See Tab. \ref{tab:Tab2} for a breakdown of publication dates and Fig. \ref{fig:Fig2} for the properties investigated.

The main contributions of this paper are:

\begin{itemize}
\item  We present various privacy preservation models for MSNs and VSNs, including, location privacy, identity privacy, anonymity, traceability, interest privacy, backward privacy, and content oriented privacy.

\item  We provide a classification for the attacks of leaking privacy for MSNs and VSNs, including, identity-based attacks, location-based attacks, eavesdropping-based attacks, manipulation-based attacks, and service-based attacks.

\item  We present various countermeasures and game theoretic approaches proposed for MSNs and VSNs.

\item  We present a side-by-side comparison in a tabular form for the current state-of-the-art of privacy-preserving schemes (thirty-three) proposed for MSNs and VSNs.

\item  We present a discussion of recommendations for further research, namely, including, privacy preserving methods, interdependent privacy, combination of privacy metrics, and identification of areas of vulnerability.
\end{itemize}

The remainder of this paper is organized as follows. Section \ref{sec:privacy-preservation-models} presents various privacy preservation models for MSNs and VSNs. In Section \ref{sec:attacks-of-leaking-privacy}, we provide a classification for the attacks of leaking privacy for MSNs and VSNs. Section \ref{sec:countermeasures-and-game-theoretic-approaches} presents various countermeasures and game theoretic approaches proposed for MSNs and VSNs. In sections \ref{sec:privacy-preserving-schemes-for-msns} and \ref{sec:privacy-preserving-schemes-for-vsns}, we present a side-by-side comparison in a tabular form for the current state-of-the-art of privacy-preserving schemes proposed for MSNs and VSNs, respectively. Finally, we discuss open issues and recommendations for further research in Section \ref{sec:recommendations-for-further-research} and draw our conclusions in Section \ref{sec:conclusion}.

\begin{table}[h]
\centering
\caption{Privacy-preserving models}
\begin{tabular}{|p{1.8in}||p{1.1in}|p{1.2in}|} \hline 
\textbf{Reference} & \textbf{Data protected} & \textbf{Model} \\ \hline 
\cite{8,50,19,26,29,30,31,32,51} & Location & Location privacy \\ \hline 
\cite{8,10,12,13,21,22,25,28,32} & ID & Identity privacy \\ \hline 
\cite{1,23,24,26,27,28,30} & ID location & Anonymity \\ \hline 
\cite{58,59,61,21,22,25} & ID & Traceability \\ \hline 
\cite{32,11} & Interest & Interest privacy \\ \hline 
\cite{28} & Backward & Backward privacy \\ \hline 
\cite{4,7,102} & All data in the network & Content oriented privacy \\ \hline 
\cite{5,11} & All data in the network & Trust evaluation \\ \hline 
\end{tabular}
\label{tab:Tab3}
\end{table}

\section{Privacy preservation models}\label{sec:privacy-preservation-models}
As shown in Tab. \ref{tab:Tab3}, the papers we review are all related with privacy preservation in ad hoc social networks, and could be divided into location privacy, identity privacy, anonymity, traceability, interest privacy, backward privacy, and content oriented privacy.

\subsection{Location privacy}

Location privacy is one of the most  important models for privacy in VSNs and MSNs, since the place of equipment (mobile phone or vehicle) can be linked to the owners. If a privacy-preserving scheme cannot guarantee the location privacy, users will be skeptical and  it cannot be accepted by the public. However, there are many solutions to ensure the location privacy in MSNs and VSNs.

In MSNs, there are two papers dealing the location privacy \cite{8,50}. Liang et al. \cite{8} proposes a proximity measurement with morality-driven data forwarding. This method provides the location privacy by mixing the hotspot based on multiplying a subgroup element. Li et al.  \cite{50} analyzes the disclosed locations \cite{2} in the MSN applications and proposes a system-level privacy control approach. This approach, via the decision tree model, it provides the privacy of location sharing in MSN.  \textbf{}

In VSNs, there are seven papers dealing the location privacy \cite{19,26} \cite{29,30,31,32,51}. Lu et al. \cite{19} proposes a strategy for the location privacy based on pseudonym self-delegated generation with conditional tracking. With this strategy, when a vehicle changes its pseudonyms, the location privacy can be guaranteed. In the paper \cite{26}, the location privacy is guaranteed using Lite-CA-based public key cryptosystem. According to Lin et al. \cite{29}, achieving receiver-location privacy preservation can be guaranteed by the social-tier dissemination phase, where the social tier request vehicles to help forwarding the packet to its neighboring social spots, and it shouldn't degrade the packet delivery performance. Lu et al.  \cite{30} proposes another technique based on pseudonyms changing at small social spot and large social spot. This technique, via anonymity analysis, can provide a promising solution for the location privacy. Related to the method proposed from Lu et al. \cite{30}, the author   proposes an idea which vehicles periodically change multiple pseudo-IDs \cite{32}. Lu et al. \cite{31} proposes another idea called ``Sacrificing the Plum Tree for the Peach Tree''. This idea is a technique based on a collection of social spots for protecting receiver-location privacy and improving the performance of packet forwarding. Yu et al. \cite{51} proposes a scheme based on a collection of social spots, including 1) global social spot and 2) individual social spot. These two types of social spot exploit the meeting opportunities for pseudonym changing in order to improve the location privacy.

\subsection{Identity privacy}

Before ensuring the location privacy, it is necessary to reassure the identity privacy. This can be acheived if in order to place of real identity, each equipment  uses a pseudonym.  There are many solutions which are based on concealing the real identity to ensure the identity privacy in MSNs and VSNs.

In MSNs, there are four papers dealing the identity privacy \cite{8,10,12,13}. Note that there are techniques that provide both identity and location privacy such as multiple-pseudonym technique \cite{8}. Lu et al. \cite{10} proposes an algorithm, called \textit{Patient Joining}, for achieving the real identity privacy by a pseudo-id. This algorithm is executed between the trusted authority and patient and it outputs a pseudo-id. Liang  et al. \cite{12} addresses the identity privacy in an emergency situation by ensuring unlinkability of the transactions and enhancing availability.  Especially, in the emergency call generation phase, the user identity is guaranteed using a group signature proposed in \cite{52}. Zhang  et al. \cite{13} develop a personalized fine-grained filtering for the identity privacy. This filtering is based on social-assisted filter distribution.

In VSNs, there are five papers dealing the identity privacy \cite{21,22,25,28,32}. Chim et al. \cite{21} support privacy-preserving of the drivers. The identity privacy is preserved using two ideas, including 1) the idea of pseudo-identity and 2) the idea of anonymous credential. Liang et al. \cite{22} proposes an idea based on a tamper-proof device activation password, which the RSU uses in order to verify the vehicle's identity and sends its tamper-proof device an anonymous credential. The trusted authority can reveal the real identity pf the vehicle to a third party for billing purposes. Chim et al. \cite{25} proposes a scheme that uses a different pseudo identity for each session. This scheme preserves the real identity based on the handshaking phase, which is executed between RSU and the trusted authority. Sun et al. \cite{28} analyzes the identity revocation based on the certificate revocation list in order to exclude an unexpired membership.  Based on the work presented in \cite{21}, Liang et al. \cite{22} presents the idea of pseudo-identity for ensuring identity privacy. Aiming to reduce the linkage between the identity and location of vehicles, Yan et al. \cite{120} propose a scheme that enhances the privacy using the idea of cell-based communication. Authors view vehicular networks as consisting of non-overlapping sub-networks restricted to a geographic area referred to as a cell. Each cell has a server that maintain a list of pseudonyms that can be assigned to the vehicles. Although the idea of dividing the network is interesting and has been extensively used for clustering  \cite{maglaras2013enhanced}  and routing reasons, the existence of a server in each area  increases the cost of the solution and also makes it infeasible. 

\subsection{Anonymity- Untraceability}

Anonymity is an important security aspect of wireless communications, since it not only protects the privacy of the users but also reduces the chances of attacks based on impersonation \cite{chen2009survey}. Untraceability is a related issue to anonymity, since if a user is traceable, its hidden identity can be revealed through the profiling of user's activity. Note that most of the anonymity schemes use a public key infrastructure (PKI) \cite{27}. For evaluating anonymity and protecting privacy, Sweeney  et al. \cite{53} proposes the k-anonymity mode. Other interesting methods which are presented in the papers \cite{54,55}, similar to the model in \cite{53}, introduce the notions of sender and receiver k-anonymity. Specifically, Wang et al. \cite{55} proposes a protocol, which ensures the anonymous transmission in a Local Ring. Moreover, there is a recent work presented in the paper \cite{1} that proposes a profile matching protocol in MSNs, called PPM, for ensuring the anonymity (from conditional to full). The PPM protocol uses three approaches, including, 1) explicit comparison-based approach, 2) implicit comparison-based approach, and 3) implicit predicate-based approach. In addition, the PPM protocol uses two anonymity enhancing techniques, including, 1) anonymity measurement and 2) anonymity enhancement.\textbf{}

In VSNs, there are six papers dealing the anonymity \cite{23,24,26,27,28,30}. Xiong et al. \cite{23} proposes a protocol that supports multi-level anonymity using the ring signature that was initialy presented in \cite{56}  and \cite{42}. Based on a pseudo-ID and anonymous certificate, Ying et al. \cite{24} proposes an idea to provide the driver with a satisfactory degree of anonymity. In the paper \cite{26}, the anonymity is guaranteed using the technique of on-path onion encryption. For generating pseudonyms, Huang et al. \cite{27} proposes an anonymity scheme, called PACP.  The proposed PACP scheme is effective and efficient compared with the anonymity schemes presented in the papers \cite{47,57} in terms of latency. Sun et al. \cite{28} proposes pseudonymous authentication scheme for the conditional anonymity, which is preserved by three techniques, including, 1) pseudonymous authentication, 2) anonymous authentication for certificate updating, and 3) certificate updating based on re-signature technology. The paper \cite{30} analysis the Quality of Privacy (QoP) with the proposal of two anonymity analytic models, including, 1) anonymity analysis on pseudonym changing at a small social spot (such as the road intersection), and 2) anonymity analysis on pseudonym changing at a large social spot (such as the free parking lot).

\subsection{Traceability}

Traceability is a very important property, where the trusted authority is able to trace a node that is misbehaving in the network. As discussed in the anonymity sub section, we have both the conditional traceability on signature and the full traceability on signature \cite{60,52}. To the best of our knowledge, there is no current work studying the traceability in MSN, but we suggest the two works presented in the papers \cite{58,59,61} for possible applicability on MSNs. Ni et al. \cite{58} proposes a protocol, called AMA, for carpooling systems. The AMA protocol supports anonymity and traceability based on the trace phase, which consists in calculating the public key of the passenger (driver) after the carpooling trip.  Sun et al. \cite{59} proposes an architecture, called SAT, for achieving the traceability based on the blind signature \cite{49,62}. The SAT architecture is an improvement of the idea presented in \cite{61}. Unlike the MSN networks, there are other recently proposed works that aim in preserving traceability in VSNs \cite{21,22,25}. Chim et al. \cite{21}, via non-repudiation property of messages, address the traceability based on the real identity of a particular vehicle, where the trusted authority can retrieve the real identity. Lu et al. \cite{22} proposes an idea based on use self-generated pseudonyms instead of real-world IDs. Chim et al. \cite{25} propose a scheme based the real identity tracking and revocation phase for satisfied the traceability and revocability. 

\subsection{Interest privacy}

Since the nodes in ad hoc social networks are formed based on a common interest, the privacy of these common interests should be preserved. The interest privacy has been explored firstly in the paper \cite{32}. The topics of common interest in the paper \cite{32} focuses on like-minded vehicles to chat. More precisely, Lu et al. in \cite{32} proposes a protocol, called FLIP, which is based on authenticated key exchange (AKE) protocols \cite{63}. Based on degree of interest verification, Rabieh et al. \cite{118} uses the attribute based encryption in order to preserve the interest privacy.

\subsection{Backward privacy}

Once the nodes in ad hoc social network have been revoked, they should reveal no information in the revocation period. There is one recently proposed work in \cite{28} that discusses the backward privacy in VSNs. Sun et al. \cite{28} propose an authentication scheme called PASS for preserving the backward privacy. The PASS scheme use the one-way hash function, which it is still difficult for any entity to reduce the certificate revocation used by the revoked vehicle. Moreover, there are some works related to the backward privacy models, such as the forward secrecy and the backward secrecy \cite{18}.

\subsection{Content oriented privacy}

As discussed in the papers \cite{4,7}, the content oriented privacy is based on three properties, including 1) immutability, 2) transparency, and 3) accountability. Ferrag et al. \cite{4} proposes a scheme called ECPDR, which is based on node certificate updating. Based an idea of certificate evolution, Ferrag et al. \cite{6} proposes a scheme called SDPP. Ferrag et al. in \cite{7} proposes another scheme called EPSA. The EPSA scheme uses the short signatures technique and the public key encryption with keyword search for ensuring content oriented privacy. There are also other works related to the content oriented privacy models, such as the impersonator resistance \cite{10,6} and the trust evaluation \cite{5,11}.

\begin{table}[h]
\centering
\caption{Summary of privacy attacks in MSNs and defense schemes}
\checkmark indicates fully supported; x indicates not supported; 0 indicates partially supported.
\begin{tabular}{|p{1.3in}||p{0.1in}|p{0.1in}|p{0.1in}|p{0.1in}|p{0.1in}|p{0.1in}|p{0.1in}|p{0.1in}|p{0.1in}|p{0.1in}|p{0.1in}|p{0.1in}|p{0.1in}|p{0.2in}|} \hline 
\begin{center}
\textbf{}
\end{center} & \multicolumn{14}{|p{2.8in}|}{\begin{center}
\textbf{Privacy-preserving schemes}
\end{center}} \\ \hline 
\textbf{Adversary model} &  \cite{1}\textbf{} &  \cite{3}\textbf{} &  \cite{4}\textbf{} &  \cite{5}\textbf{} &  \cite{6}\textbf{} &  \cite{7}\textbf{} &  \cite{8}\textbf{} &  \cite{9}\textbf{} & \cite{10}\textbf{} & \cite{11}\textbf{} & \cite{12}\textbf{} & \cite{13}\textbf{} & \cite{50}\textbf{} & \cite{102} \\ \hline 
\textbf{Forgery attack\newline } & \checkmark & \checkmark & x & x & x & x & x & \checkmark & x & x & \checkmark & 0 & x & x \\ \hline 
\textbf{Attribute-trace attack\newline } & x & x & 0 & x & 0 & 0 & x & \checkmark & x & x & x & x & 0 & 0 \\ \hline 
\textbf{Eavesdropping attack\newline } & x & x & 0 & x & 0 & 0 & x & \checkmark & x & x & x & x & 0 & 0 \\ \hline 
\textbf{Collusion attack\newline } & x & x & x & x & x & x & x & \checkmark & 0 & x & \checkmark & 0 & x & x \\ \hline 
\textbf{Wormhole attack\newline } & x & x & 0 & x & 0 & \checkmark & 0 & 0 & x & x & x & x & 0 & 0 \\ \hline 
\textbf{Black hole attack\newline } & x & x & \checkmark & x & \checkmark & 0 & x & 0 & x & x & x & x & 0 & x \\ \hline 
\textbf{Sybil attack\newline } & x & x & x & \checkmark & x & x & x & x & x & x & x & x & x & x \\ \hline 
\textbf{Chosen-plaintext attack\newline } & x & x & x & x & x & x & 0 & 0 & \checkmark & x & 0 & 0 & x & 0 \\ \hline 
\textbf{Spam attack\newline } & x & x & x & x & x & x & x & x & x & \checkmark & x & x & x & x \\ \hline 
\textbf{Identity theft attack\newline } & 0 & 0 & x & x & x  & x & x & 0 & x & x & \checkmark & 0 & x & x \\ \hline 
\textbf{User manipulation attack\newline } & x & x & x & x & x & x & \checkmark & 0 & 0 & x & 0 & 0 & x & x \\ \hline 
\textbf{Message tampering attack\newline } & x & \checkmark & 0 & x & 0 & 0 & x & 0 & x & x & x & x & 0 & x \\ \hline 
\textbf{Routing attack\newline } & x & x & \checkmark & x & \checkmark & 0 & x & 0 & x & x & x & x & 0 & x \\ \hline 
\textbf{Linkability attack\newline } & x & x & 0 & \checkmark & 0 & 0 & x & 0 & x & x & x & x & 0 & x \\ \hline 
\textbf{Rejection attack \newline } & x & x & 0 & \checkmark & 0 & 0 & x & 0 & x & x & x & x & 0 & x \\ \hline 
\textbf{Modification attack\newline } & x & x & 0 & \checkmark & 0 & 0 & x & 0 & x & x & x & x & 0 & x \\ \hline 
\textbf{Inside curious attack\newline } & x & x & x & x & x & x & 0 & 0 & 0 & x & 0 & \checkmark & x & 0 \\ \hline 
\textbf{Outside forgery attack\newline } & 0 & 0 & x & x & x & x & x & 0 & x & x & 0 & \checkmark & x & x \\ \hline 
\textbf{Demographic inference attack} & x & x & 0 & x & 0 & 0 & x & 0 & x & x & x & x & \checkmark & x \\ \hline 
\textbf{User-profiling attack\newline } & x & x & x & x & x & x & x & 0 & x & x & x & 0 & x & \checkmark \\ \hline 
\end{tabular}\\
\label{tab:Tab4}
\end{table}
\begin{table}[h]
\centering
\caption{Summary of privacy attacks in VSNs and defense schemes}
\checkmark indicates fully supported; x indicates not supported; 0 indicates partially supported.
\begin{tabular}{|p{1.2in}||p{0.1in}|p{0.1in}|p{0.1in}|p{0.1in}|p{0.1in}|p{0.1in}|p{0.1in}|p{0.1in}|p{0.1in}|p{0.1in}|p{0.1in}|p{0.1in}|p{0.1in}|p{0.1in}|p{0.1in}|p{0.1in}|p{0.1in}|p{0.1in}|} \hline 
\textbf{} & \multicolumn{18}{|p{4.2in}|}{\textbf{\begin{center}
Privacy-preserving schemes
\end{center}}} \\ \hline 
\textbf{Adversary model} &  \cite{18} & \cite{19} & \cite{20} & \cite{21}\textbf{} & \cite{22}\textbf{} & \cite{23}\textbf{} & \cite{24}\textbf{} &  \cite{25}\textbf{} & \cite{26}\textbf{} & \cite{27}\textbf{} & \cite{28}\textbf{} & \cite{29}\textbf{} & \cite{30}\textbf{} &  \cite{31}\textbf{\newline } & \cite{32}\textbf{\newline } & \cite{33}\textbf{\newline } & \cite{34}\textbf{\newline } &  \cite{51}\textbf{} \\ \hline 
\textbf{Sybil attack\newline } & \checkmark & x & \checkmark & x & \checkmark & 0 & x & 0 & x & 0 & x & 0 & x & x & x & x & 0 & 0 \\ \hline 
\textbf{Global external attack\newline } & x & \checkmark & x & x & 0 & x & 0 & x & x & x & x & x & x & x & x & x & x & 0 \\ \hline 
\textbf{Replaying attack\newline } & x & x & \checkmark & \checkmark & x & \checkmark & x & \checkmark & x & \checkmark & \checkmark & x & x & x & x & x & \checkmark & \checkmark \\ \hline 
\textbf{Impersonation attack\newline } & x & x & 0 & x & 0 & \checkmark & x & \checkmark & x & \checkmark & x & \checkmark & x & x & x & x & \checkmark & \checkmark \\ \hline 
\textbf{Location tracking attack\newline } & x & 0 & x & x & \checkmark & x & 0 & x & x & x & x & x & x & x & x & x & x & 0 \\ \hline 
\textbf{Identity revealing attack\newline } & 0 & x & 0 & x & \checkmark & 0 & x & 0 & x & 0 & x & 0 & x & x & x & x & 0 & 0 \\ \hline 
\textbf{Eavesdropping attack\newline } & x & x & x & x & x & x & \checkmark & x & x & \checkmark & x & x & \checkmark & 0 & 0 & 0 & \checkmark & 0 \\ \hline 
\textbf{Forgery attack\newline } & x & 0 & x & x & 0 & x & \checkmark & x & x & x & x & x & x & x & x & x & x & 0 \\ \hline 
\textbf{Chosen plaintext attack\newline } & x & x & x & x & x & x & x & x & \checkmark & x & x & x & x & x & x & x & x & 0 \\ \hline 
\textbf{Chosen ciphertext attack\newline } & x & x & x & x & x & x & x & x & \checkmark & x & x & x & x & x & x & x & x & 0 \\ \hline 
\textbf{Adaptive chosen ciphertext attack} & x & x & x & x & x & x & x & x & \checkmark & x & x & x & x & x & x & x & x & 0 \\ \hline 
\textbf{Man-in-the-middle attack} & x & x & x & x & x & x & x & x & \checkmark & x & x & x & x & x & x & x & x & 0 \\ \hline 
\textbf{Denial-of-service (DoS) attack} & x & x & 0 & 0 & x & 0 & x & 0 & x & 0 & \checkmark & x & x & x & x & x & 0 & 0 \\ \hline 
\textbf{Source bogus attack\newline } & x & x & 0 & 0 & x & 0 & x & 0 & x & 0 & 0 & \checkmark & x & x & x & x & 0 & 0 \\ \hline 
\textbf{Black/grey hole attack\newline } & x & x & x & x & x & x & 0 & x & x & 0 & x & \checkmark & 0 & \checkmark & 0 & \checkmark & 0 & x \\ \hline 
\textbf{Successive-response attack} & x & x & x & x & x & x & 0 & x & x & 0 & x & x & 0 & 0 & \checkmark & 0 & 0 & x \\ \hline 
\textbf{Packet analysis attack\newline } & x & x & x & x & x & x & 0 & x & x & 0 & x & x & 0 & 0 & 0 & \checkmark & 0 & x \\ \hline 
\textbf{Packet tracing attack\newline } & x & x & x & x & x & x & 0 & x & x & 0 & x & x & 0 & 0 & 0 & \checkmark & 0 & x \\ \hline 
\textbf{Brute force cryptanalytic attack} & x & x & x & x & x & x & x & x & 0 & x & x & x & x & x & x & x & x & \checkmark \\ \hline 
\textbf{Incorrect data attack\newline } & x & x & x & x & x & x & x & x & 0 & x & x & x & x & x & x & x & x & \checkmark \\ \hline 
\textbf{Liability attack\newline } & x & x & x & x & x\newline  & x & x & x & 0 & x & x & x & x & x & x & x & x & \checkmark \\ \hline 
\end{tabular}
\label{tab:Tab5}
\end{table}
\begin{figure}[h]
\centering
\includegraphics[width=1\linewidth]{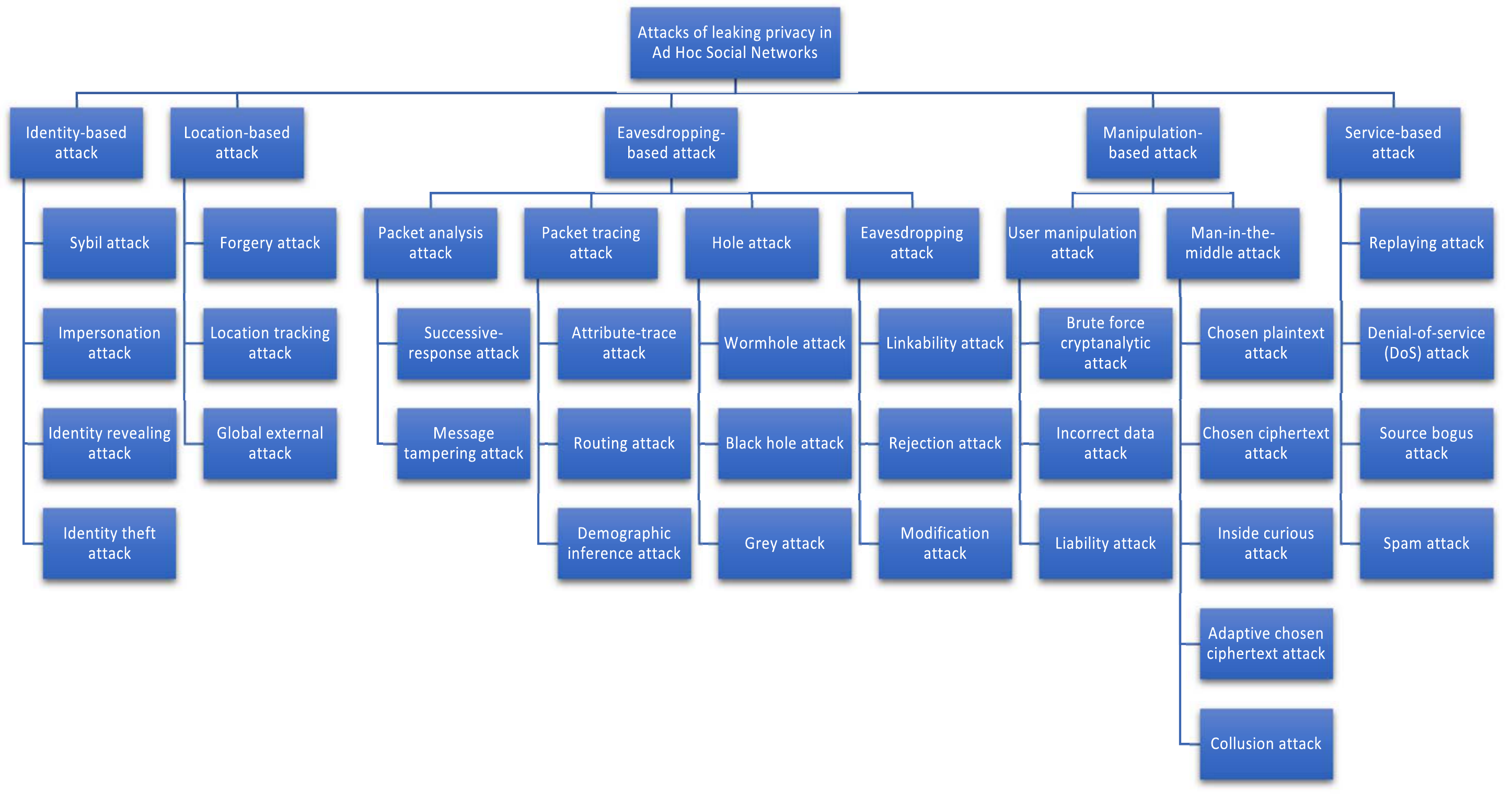}
\caption{Classification of attacks of leaking privacy in Ad Hoc Social Networks}
\label{fig:Fig3}
\end{figure}

\section{Attacks of leaking privacy}\label{sec:attacks-of-leaking-privacy}

In this section, we discuss the attacks of leaking privacy in Ad Hoc Social Networks. The classification of attacks in ad hoc networks frequently mentioned in literature is done using different criteria such as passive or active, internal or external \cite{65,66,67,68} etc. In our survey articvle  we classify the attacks of leaking privacy in five categories as shown in Fig. \ref{fig:Fig3}, including, 1) identity-based attack, 2) location-based attack, 3) eavesdropping-based attack, 4) manipulation-based attack, and 5) service-based attack. In addition, Tab. \ref{tab:Tab4} and Tab. \ref{tab:Tab5} give a detailed summary of security threats in MSNs and VSNs, respectively.

\subsection{Identity-based attack}

The attacks that belong to this category are somehow related to the manipulation of the identity of legitimate users. As identity based attacks we can characterize the Sybil attack and the Impersonation attack.

\begin{itemize}
\item  \textbf{Sybil attack: }When an adversary node has multiple identities, a Sybil attack can be launched in the ad hoc social network. The major goal of the adversary in this attack is to be a destination repeatedly. Once the packets are routed to him, then he can realize other types of attacks such as the selective forwarding attack \cite{64}. The TSE system \cite{5} can resist the sybil attacks using the idea of multiple reviews in a short time period. The scheme in \cite{20} can post detected the Sybil attacks by multiple valid pseudo-IDs. The framework in \cite{22} can detect the Sybil attacks using the ID-based signature and the ID-based online/offline signature.

\item  \textbf{Impersonation attack: }During the registration phase, when the vehicle with real identity generates its pseudo identity, an impersonation adversary records this pseudo identity, which he can realize other types of attacks such as the identity revealing attack \cite{22} and the identity theft attack \cite{12}. The protocol in \cite{23}, the PACP protocol in \cite{27}, and the SECSPP scheme in \cite{34} use authentication of messages to guard against impersonation attacks. The SPECS schemes in \cite{25} uses a phase called real identity tracking and revocation against an impersonation adversary. Since the single-attribute encryption is employed in the STAP protocol \cite{29}, an impersonation adversary can be detected. With the use the safety message and include a valid certificate from the trusted register authority, the MixGroup scheme in \cite{51} can avoid the impersonation attacks.\textbf{}
\end{itemize}

\subsection{Location-based attack}

This category of attacks is based on revealing the user location, and it consists of two major attacks the forgery attack and the global external attack.

\begin{itemize}
\item  \textbf{Forgery attack: }In this attack, a forgery adversary generates a misleading message with the bogus location information in order to start the plotting attacks as the location tracking attack \cite{22}. Using an explicit comparison-based approach, the PPM protocol in \cite{1} has been proven with a theorem (non-forgeability) that any profile forgery attack can be detected. The SFPM protocol in \cite{3} can resist to forgery attacks using a data processing center. The HealthShare scheme in \cite{9} can be effectively resisted to the forgery attacks using the attribute-oriented authentication scheme. The PEC scheme in \cite{12} can withstand to the forgery attacks via the group signature, which helps the trusted authority to track the user's unique identity. The PPBMA scheme in \cite{24} can prevent the forgery attacks based on the verification of this equation,  ${MAC}_{k^i_j}(M^F_j||T^F_j)=?{MAC}^{Fi}_{j+1}(M^F_j||T^F_j)$. 

\item  \textbf{Global external attack: }This attack is proposed in the paper \cite{19}, which can be classified in this category, i.e., location-based attack. More precisely, a global external adversary tracks a vehicle in terms of Time, Location, and Velocity. The PCS strategy in \cite{19} can resist to the global external attacks using two phases, including, 1) pseudonym self-delegated generation, and 2) conditional tracking.\textbf{}
\end{itemize}

\subsection{Eavesdropping-based attack}

This category of attacks is based on eavesdropping the network communications, and it consists of four major attacks; 1) eavesdropping attack, 2) packet analysis attack, 3) packet tracing attack, and 4) hole attack.

\begin{itemize}
\item  \textbf{Eavesdropping attack: }When the nodes in ad hoc social network try to exchange the information on common interests, an eavesdropping attack tries to attain the transmitted data without the certificates. Then, it can perform some operations on this data using the linkability attack, rejection attack, and modification attack \cite{3}. The HealthShare scheme in \cite{9} is resistant to the eavesdropping attacks based on the ciphertext generated by delegated encryption algorithm. The PPBMA scheme in \cite{24} is resistant against the eavesdropping attacks using the following condition: ${MAC}_{k^i_j}(M_j||T_j)\ne {MAC}_{k^{i'}_j}(M_j||T_j)$. Based on semantic security, the PACP protocol in \cite{27} has been proved that is semantically secure against the eavesdropping attacks. The SECSPP scheme in \cite{34} has been proved that is secured against the eavesdropping attacks based on the authorization access phase.

\item \textbf{Packet analysis attack: }This is a type of attack that is popular in wired networks, where an adversary captures the packets, and then it analyses in order to extract important information such as the common interests. In ad hoc social networks, this attack has the same strategic but the probability of launch is high compared to wired networks \cite{33}. In addition, the successive-response attack \cite{32} and the message tampering attack \cite{3} could be run through the packet analysis attack.  However, the SPRING protocol in \cite{33} can resist to the packet analysis attacks using the anonymous authentication, which is based on a conditional privacy-preserving authentication technique. This technique is based on two key phases, including, 1) privacy-preserving authentication, and 2) conditional tracking.
\item  \textbf{Packet tracing attack: }After eavesdropping the source and destination locations of packet, an adversary can trace this packet without need to recover the packet content \cite{33}. In addition, when a security scheme uses the attributes, an attack can be launched, called the attribute-trace attack \cite{9}. Therefore, the routing attack \cite{4,6} and demographic inference attack \cite{50} could be run by tracing the packet controls used by the routing protocol.  The SPRING protocol in \cite{33} can resist to the packet tracing attacks based on the anonymous authentication.

\begin{figure}[h]
\centering
\includegraphics[width=0.5\linewidth]{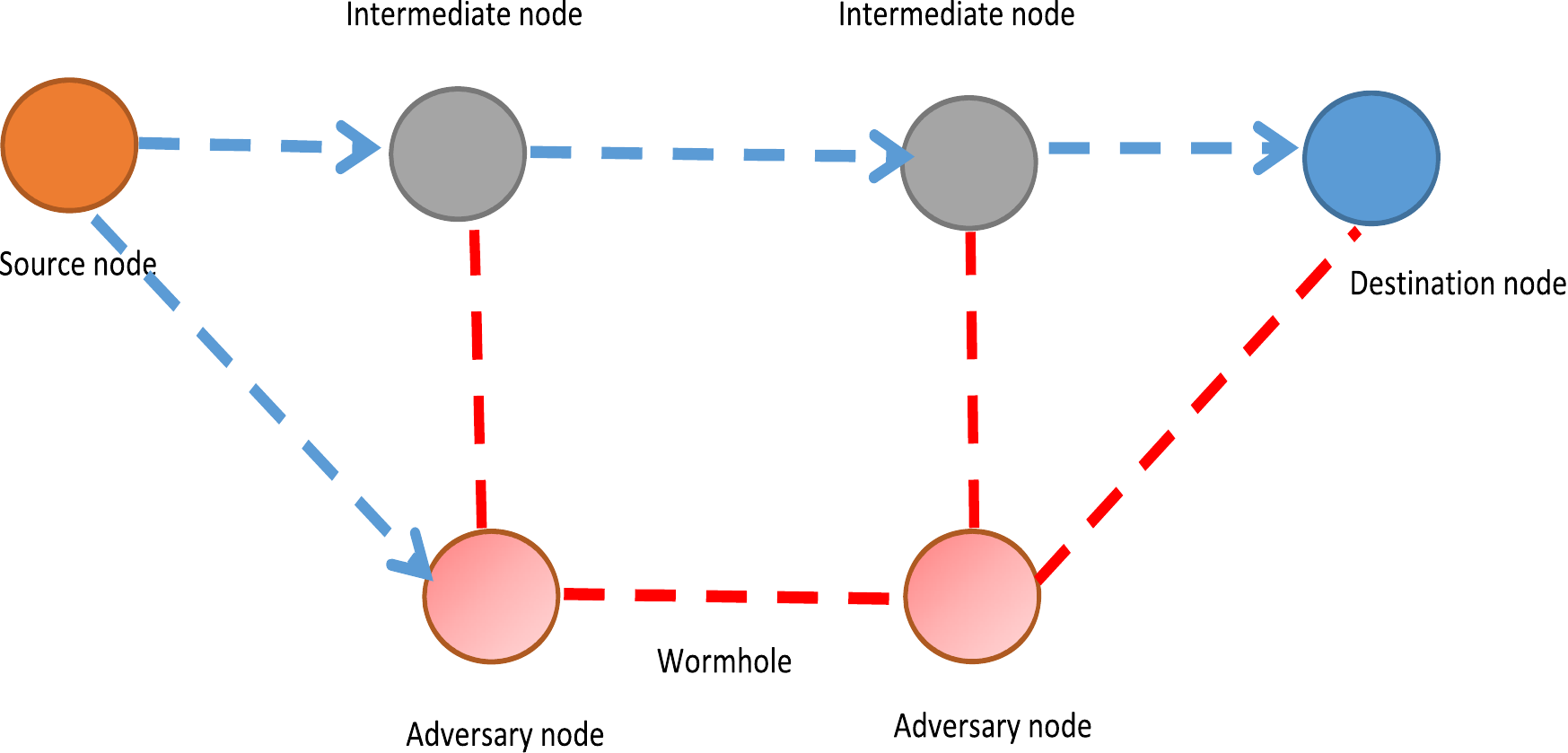}
\caption{Wormhole attack}
\label{fig:Fig4}
\end{figure}

\begin{figure}[h]
\centering
\includegraphics[width=0.4\linewidth]{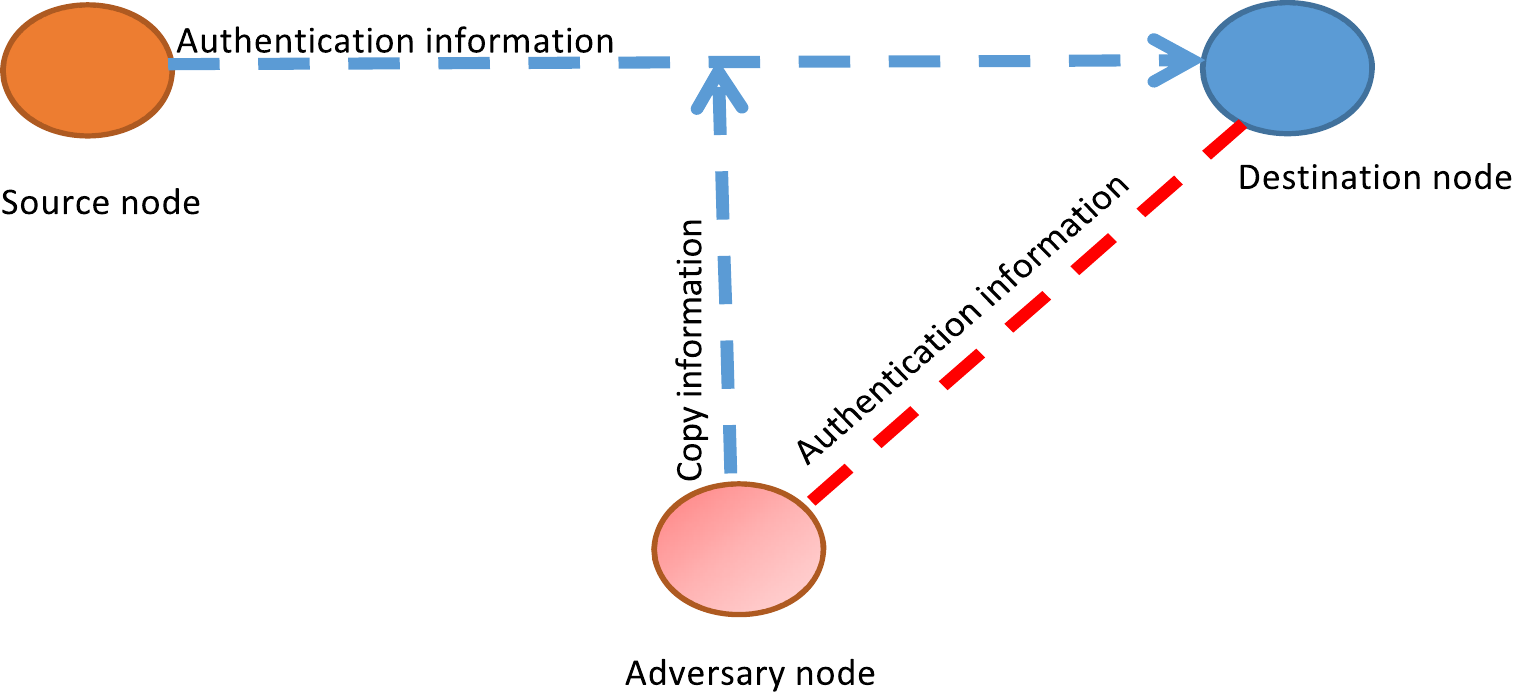}
\caption{Replay attack}
\label{fig:Fig5}
\end{figure}

\begin{table}[h]
\centering
\caption{Approaches for detecting and avoiding the hole attacks}
\begin{tabular}{|p{0.7in}||p{0.8in}|p{0.8in}|p{1.2in}|} \hline 
\textbf{Scheme} & \textbf{Type} & \textbf{Data attacked} & \textbf{Approach} \\ \hline 
SPRING \cite{33}\newline (2010) & Black/grey hole  & Sensitive data & Conditional authentication technique \\ \hline 
SDPP \cite{6}\newline (2014) & Black hole & Proactive Routing data & Cooperative neighbor X neighbor \\ \hline 
STAP \cite{29}\newline (2011) & Black/grey hole  & Sensitive data & Signature algorithm \\ \hline 
ECPDR \cite{4}\newline (2013) & Wormhole attack & User data & Proxy re-signature cryptography \\ \hline 
SPF \cite{31}\newline (2010) & Black/grey hole  & Sensitive data & Anonymous identity-based encryption \\ \hline 
EPSA \cite{7} \newline (2016) & Wormhole attack & Reactive Routing data & Cooperative neighbor X neighbor \\ \hline 
\end{tabular}
\label{tab:Tab6}
\end{table}

\begin{table}[h]
\centering
\caption{Approaches for detecting and avoiding the replay attacks}
\begin{tabular}{|p{0.8in}||p{1.7in}|p{1.1in}|} \hline 
\textbf{Reference} & \textbf{Data attacked} & \textbf{Approach} \\ \hline 
Lu et al. \cite{20}\newline (2010) & Inner data stored in the OBU & Checks $|T'-T|\ \le \ \Delta T$ \\ \hline 
Chim et al. \cite{21}\newline (2014) & Local data of RSU & Timestamps \\ \hline 
Xiong et al. \cite{23}\newline (2012) & Data of member manager & Timestamps \\ \hline 
Chim et al. \cite{25}\newline (2011) & Data broadcast by the RSU & Timestamps \\ \hline 
Huang et al. \cite{27}\newline (2011) & All data in the network & Timestamps \\ \hline 
Sun et al. \cite{28}\newline (2010) & Data broadcast by the vehicles & Timestamps \\ \hline 
Li et al. \cite{34}\newline (2008) & Data transmitted between the vehicles & Sequence numbers \\ \hline 
Yu et al. \cite{51}\newline (2016) & Data broadcast by the vehicles & Timestamps \\ \hline 
\end{tabular}
\label{tab:Tab7}
\end{table}

\item  \textbf{Hole attack: }This category consists of three types of attacks, including, wormhole attack, black hole attack, and grey hole attack. The hole attack is based on creating a communication tunnel where an adversary eavesdrops the communication inside the ad hoc social network through this tunnel, as shown in Fig. \ref{fig:Fig4}. Note that several adversaries can initiate the creation of the tunnel, where the routing protocol will be totally under the control of these adversaries. Tab \ref{tab:Tab6}. shows the approaches for detecting and avoiding the hole attacks in ad hoc social networks. For detecting  blackhole attacks, the SDPP scheme proposed in \cite{6} usse the cooperative neighbor technique and the homomorphic encryption method. The ECPDR scheme in \cite{4} can resist to a wormhole attack using a restore strategy with the proxy re-signature cryptography technology. The EPSA scheme in \cite{7} can detect and prevent a wormhole attack  by using the cooperative neighbor X neighbor, its performance depending on the length of the tunnel created from the  adversaries, i.e., the longer the tunnel,  the higher the detection rate. Tracking the inside black/grey hole adversaries is possible with the STAP protocol which is presented in \cite{29} with the use of the validity of $sig(CTL)$, where $sig$ is a signature algorithm and $CTL$ is the timestamp/location information. The SPF protocol \cite{31} can resist to black/grey hole attacks using the anonymous identity-based encryption. The SPRING protocol \cite{33} can also track inside black/grey hole adversaries by using the conditional privacy-preserving authentication technique. In addition, the SPRING protocol can detect  black/grey hole attacks with a detection algorithm, which is based on the distance $d(X_i)$ of each node $X_i$ in all vehicle nodes $V$ to the mean $\overline{X}$ and the thresholds $T_B$, $T_G$ for black hole attack and grey hole attack, where $d(X_i)\ =\ |X_i-\ \overline{X}|$, $\overline{X}=\frac{1}{\left| V\right|}\sum^{\left|V\right|}_{i=1}{X_i}$. The node is considered as a grey/black hole adversary when $d\left(X_i\right)>T_{{\mathbf G}}$ or $d\left(X_i\right)>T_B$.
\end{itemize}

\subsection{Manipulation-based attack}

This category of attacks is based on the manipulation of nodes of the Ad Hoc Social Network (users or socialspots), and it consists of two major attacks; 1) user manipulation attack and 2) man-in-the-middle attack.

\begin{itemize}
\item  \textbf{User manipulation attack: }In this attack, an adversary tries to appear as a hotspot (small or large) for the nodes in ad hoc social network, in order to have the updates certificates. For example, an adversary sends a packet containing information of a false hotspot, then the node run the update certificates phase with this adversary. Therefore, these nodes have to honestly tell about their hotspots \cite{8}. In addition, the brute force cryptanalytic attack, incorrect data attack, and liability attack \cite{51}, could be run through the user manipulation attack. The protocol proposed in \cite{8} uses the authentication against the user manipulation attack. Specifically, this protocol is based on a privacy-preserving routing tree in order to make sensitive hotspots anonymous.

\item  \textbf{Man-in-the-middle attack:} The idea of this attack is like the idea in the hole attacks, but we classify it in this category because the adversary manipulates the users. The man-in-the-middle attack is especially applicable in the Diffie--Hellman key exchange method. As discussed in the survey \cite{117}, man-in-the-middle attack aims to compromise confidentiality, integrity, and availability. However, in order to reduce the security of the encryption scheme, the adversary can launch other types of attacks in this category as the chosen plaintext attack, chosen ciphertext attack, inside curious attack, and adaptive chosen ciphertext attack \cite{26}. In addition, an adversary can launch a collusion attack \cite{9,12} where he tries to find two different packets $p1$ and $p2$ such that $hash(p1)\ =\ hash(p2)$. The EP2DF scheme in \cite{26} proposed a novel authentication framework, called lite-CA-based public key cryptosystem, to thwart the man-in-the-middle attacks and has been proved that is secure against adaptive chosen ciphertext attacks. Recall that the schemes resist against the eavesdropping attacks can resist against the collusion attacks. Based on the independent relation of the secret shares, the PEC scheme in \cite{12} can avoid the collusion attacks.
\end{itemize}

\subsection{Service-based attack}

This category of attacks aiming to make a service unavailable of the network and it consists of four major attacks; 1) replaying attack, 2) denial-of-service (DoS) attack, 3) source bogus attack, and 4) spam attack.

\begin{itemize}
\item  \textbf{Replaying attack: }When a node A wants to exchange data with node B, the node A must prove its identity, which the node B request a valid certificate, i.e., authentication information as shown in Fig. \ref{fig:Fig5}. Then, the node A sent this certificate in a signed packet (SP). During this exchange, an adversary listening and saves this signed packet. Once the exchange is completed, this adversary tries to contact the node B. Hence, the node B request a valid certificate to adversary. The adversary sent the SP to the node B. At the end, the node B believes be dealing with the node A, and a service can be unavailable by the adversary. Note that this attack was discussed in several papers \cite{20,21,23,25,27,28,34,51}. Tab \ref{tab:Tab7}. shows the approaches for detecting and avoiding the replay attacks in ad hoc social networks. After the verification of the identity $PIDi$, by a valid time interval $\Delta T$ for transmission delay, the intelligent parking scheme in \cite{20} can avoid the replaying attacks when the RSU checks the following condition : $|T'-T|\ \le \ \Delta T$. Similarly to the scheme in \cite{20}, the scheme in \cite{21}, the scheme in \cite{23}, the PASS scheme in \cite{28}, and the SECSPP scheme in \cite{34} checks the timestamps in the messages to reduce the impact of replay attack. When the RSU stores the pseudo-identities used by vehicles, the SPECS scheme in \cite{25} can avoid the replay attacks with the help of RSU, which he can checks the pseudo-identity in its database. The PACP protocol in \cite{27} can avoid the replay attacks with the use of authentication and sequence numbers. The MixGroup scheme in \cite{51} can avoid the replay attacks with the use of timestamps in the revocation operation.

\item  \textbf{Denial-of-service (DoS) attack: }This attack is the heart of this category. During the social communications, an adversary can launch a DoS attack in order to put a service unavailable, for example, disruption of routing process, block a file server, wasting the limited buffer resource, or preventing the distribution of secret keys.  Therefore, DoS attack can be launched from several layers, i.e., link layer, physical layer, network layer, transport layers, and application layers \cite{65}. To detect the Denial-of-service (DoS) attacks, the PASS scheme in \cite{28} adopt the Schnorr signature algorithm and the prestore strategy to signing certificate ${{\rm Cert}}_{TA,V_i,k}$ and pseudonymous certificate ${{\rm Cert}}_{TA,V_i,j}$. 

\item  \textbf{Source bogus attack: }This attack is qualified as an inside attack, which a source deliberately inject bogus data in the social communications to wasting the limited buffer resource of the nodes \cite{29}. In addition, this attack is similar to the incorrect data attack \cite{51} but with other objectives. Note that we found only Lin et al. \cite{29} that discuss the source bogus attack in the ad hoc social networks until now. The STAP protocol in \cite{29} uses the single-attribute encryption against the source bogus attacks.

\item  \textbf{Spam attack: }This attack is very popular in electronic mails. In general, an adversary tries to send several e-mails to multiple receivers whose addresses have generally been recovered on the internet. The first goal of this attack is to do advertising at lower prices. However, an adversary can launch this attack in ad hoc social networks in order firstly to disrupt the data filtering and secondly to spy the storage space. Note that we found only Hameed et al. \cite{11} that discuss the spam attack in the ad hoc social networks until now. The LENS system in \cite{11} can prevent the spam transmission using the idea of Gate keepers.
\end{itemize}
\begin{table}[h]
\centering
\caption{Cryptographic methods used in privacy-preserving schemes for MSNs}
\checkmark indicates that the scheme uses the cryptographic method.
\begin{tabular}{|p{1.9in}||p{0.1in}|p{0.1in}|p{0.1in}|p{0.1in}|p{0.1in}|p{0.1in}|p{0.1in}|p{0.1in}|p{0.1in}|p{0.1in}|p{0.1in}|p{0.1in}|} \hline 
\textbf{} & \multicolumn{12}{|p{3.2in}|}{\textbf{Privacy-preserving schemes}} \\ \hline 
\textbf{Cryptographic methods} &  \cite{1} & \cite{3} &  \cite{4} &  \cite{5} &  \cite{6} &  \cite{7} &  \cite{8} &  \cite{9} &  \cite{10} &  \cite{12} &  \cite{13} & \cite{102} \\ \hline 
Secure cryptographic hash functions \cite{36} &  &  & \checkmark & \checkmark & \checkmark & \checkmark & \checkmark & \checkmark & \checkmark & \checkmark & \checkmark &  \\ \hline 
Homomorphic encryption \cite{85} & \checkmark &  &  &  & \checkmark &  &  &  &  &  &  & \checkmark \\ \hline 
Multiple pseudonym technique \cite{86} &  &  &  &  &  &  & \checkmark &  &  &  &  &  \\ \hline 
HMAC \cite{87} &  & \checkmark &  &  &  &  &  &  &  &  &  &  \\ \hline 
Short signatures technique \cite{37} &  &  & \checkmark &  &  & \checkmark &  &  &  &  &  &  \\ \hline 
Identity-Based Aggregate Signatures \cite{88} &  &  &  & \checkmark &  &  &  &  &  &  &  &  \\ \hline 
Identity-based encryption \cite{38} &  &  &  &  & \checkmark &  &  &  &  &  &  &  \\ \hline 
PKE with keyword search \cite{89} \cite{90} &  &  &  &  &  & \checkmark &  &  &  &  &  &  \\ \hline 
Linear Secret Sharing \cite{91} &  &  &  &  &  &  &  & \checkmark &  & \checkmark &  &  \\ \hline 
Attribute-based encryption \cite{93} &  &  &  &  &  &  &  &  &  & \checkmark &  &  \\ \hline 
Hidden vector encryption \cite{94}\cite{97} &  &  &  &  &  &  &  &  &  &  & \checkmark &  \\ \hline 
Short group signature \cite{52} &  &  &  &  &  &  &  &  &  & \checkmark &  &  \\ \hline 
\end{tabular}
\label{tab:Tab8}
\end{table}

\begin{table}[h]
\centering
\caption{Cryptographic methods used in privacy-preserving schemes for VSNs}
\checkmark indicates that the scheme uses the cryptographic method.
\begin{tabular}{|p{1.9in}||p{0.1in}|p{0.1in}|p{0.1in}|p{0.1in}|p{0.1in}|p{0.1in}|p{0.1in}|p{0.1in}|p{0.1in}|p{0.1in}|p{0.1in}|p{0.1in}|p{0.1in}|p{0.1in}|p{0.1in}|p{0.1in}|p{0.1in}|} \hline 
\textbf{} & \multicolumn{17}{|p{3in}|}{\textbf{Privacy-preserving schemes}} \\ \hline 
\textbf{Cryptographic methods} & \cite{18} & \cite{19} & \cite{20} & \cite{21} &  \cite{22} & \cite{23} & \cite{24} & \cite{25} & \cite{26} & \cite{27} & \cite{28} & \cite{29} & \cite{31} & \cite{32} & \cite{33} & \cite{34} & \cite{51} \\ \hline 
Secure cryptographic hash functions \cite{36} & \checkmark & \checkmark &  & \checkmark &  & \checkmark & \checkmark & \checkmark & \checkmark & \checkmark & \checkmark & \checkmark & \checkmark & \checkmark & \checkmark & \checkmark & \checkmark \\ \hline 
Verifier-local revocation \cite{35} & \checkmark &  &  &  &  &  &  &  &  &  &  &  &  &  &  &  & \checkmark \\ \hline 
Short signatures technique \cite{37} &  & \checkmark &  & \checkmark &  &  &  & \checkmark &  & \checkmark &  &  &  &  &  &  & \checkmark \\ \hline 
Ephemeral key \cite{92} &  &  & \checkmark &  &  &  &  &  &  &  &  &  &  &  &  &  &  \\ \hline 
Identity-based encryption \cite{38} &  &  & \checkmark &  &  &  &  &  &  & \checkmark &  &  &  &  &  &  &  \\ \hline 
Proxy Re-encryption \cite{39} &  &  &  & \checkmark &  &  &  &  &  &  &  &  &  &  &  &  &  \\ \hline 
ID-based signature \cite{40} &  &  &  &  & \checkmark &  &  &  &  &  &  &  &  &  &  &  &  \\ \hline 
ID-based online/offline signature \cite{41} &  &  &  &  & \checkmark &  &  &  &  &  &  &  &  &  &  &  &  \\ \hline 
Ring signature \cite{42} &  &  &  &  &  & \checkmark &  &  &  &  &  &  &  &  &  &  &  \\ \hline 
Certificateless public key \cite{43}\cite{44} &  &  &  &  &  &  &  &  & \checkmark &  &  &  &  &  &  &  &  \\ \hline 
Schnorr signature algorithm \cite{45} &  &  &  &  &  &  &  &  &  &  & \checkmark &  &  &  &  &  &  \\ \hline 
Elliptic curve (ECDSA) \cite{46} &  &  &  &  &  &  &  &  &  &  &  &  &  & v &  &  &  \\ \hline 
CPPA Technique \cite{47} &  &  &  &  &  &  &  &  &  &  &  &  &  &  & \checkmark &  &  \\ \hline 
Blind signature  \cite{49}\cite{62} &  &  &  &  &  &  &  &  &  &  &  &  &  &  &  & \checkmark &  \\ \hline 
Non-interactive ID-based PKC \cite{48} &  &  &  &  &  &  &  &  &  &  &  &  &  &  &  & \checkmark &  \\ \hline 
Multiple pseudonym technique \cite{86} &  &  &  &  &  &  &  &  &  &  &  &  &  &  &  &  & \checkmark \\ \hline 
\end{tabular}
\label{tab:Tab9}
\end{table}

\begin{figure}[h]
\centering
\includegraphics[width=0.6\linewidth]{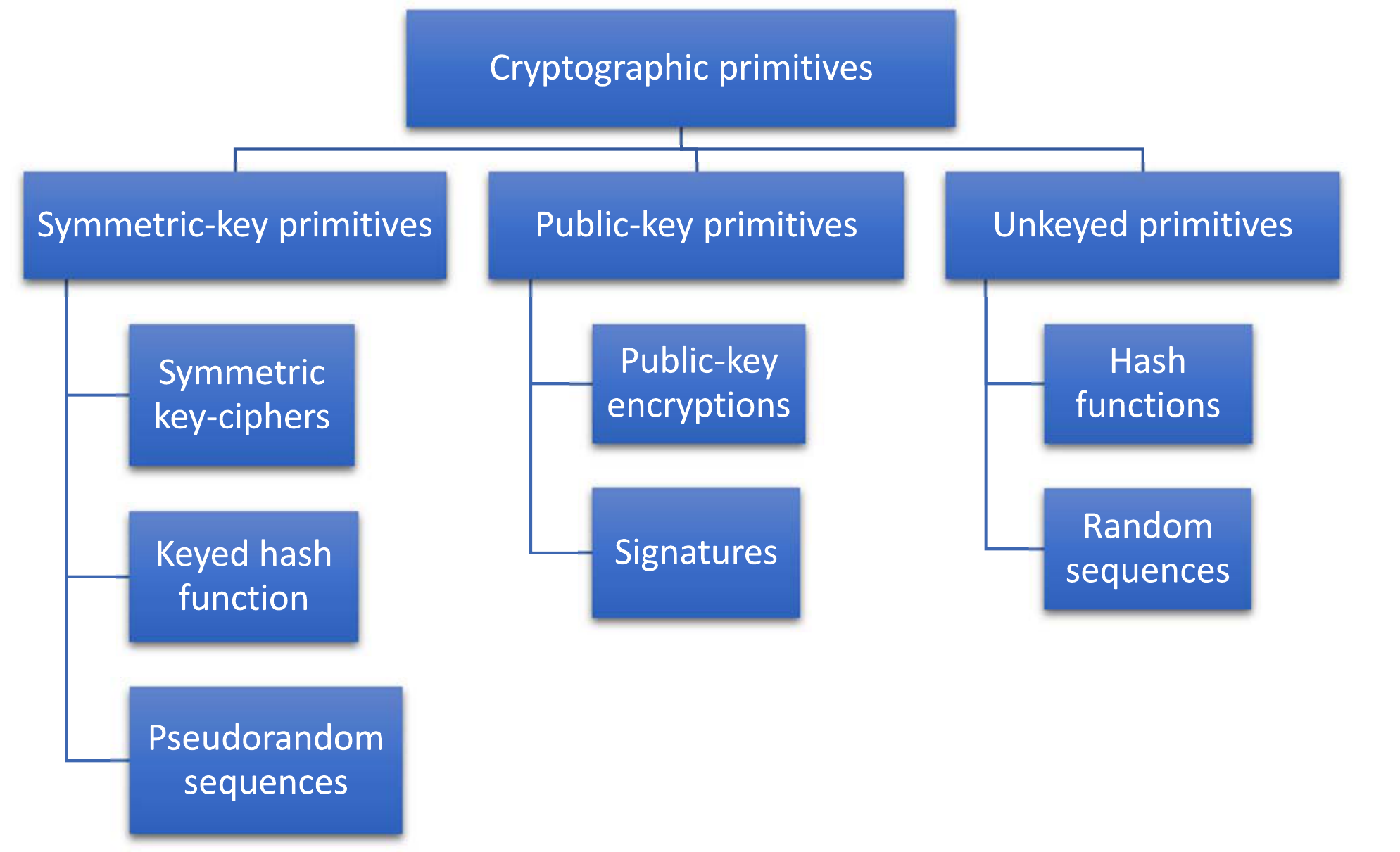}
\caption{Taxonomy of cryptographic primitives}
\label{fig:Fig6}
\end{figure}

\section{Countermeasures and game theoretic approaches}\label{sec:countermeasures-and-game-theoretic-approaches}
Most privacy preserving schemes for ad hoc social networks that we have examined use the cryptography as a countermeasure in order to preserve  privacy.  Generally, cryptography is one of the disciplines of cryptology, which was initialy  proposed in order to protect data, i.e., ensuring confidentiality, authenticity, and integrity using secrets or keys. Fig. \ref{fig:Fig6} presents the taxonomy of cryptographic primitives \cite{82}. The cryptographic methods used in privacy-preserving schemes for MSNs and VSNs are summarized in Tab. \ref{tab:Tab8} and Tab. \ref{tab:Tab9}, respectively. In order to prove these security schemes theoretically, researchers can use game theoretic approaches \cite{83,84}. In this section, we will discuss the state of the art of privacy-preserving schemes for VSNs and MSNs and we provide a description and types of game theoretic models for preserving privacy in ad hoc social networks.

\subsection{Symmetric-key primitives}
Symmetric-key primitives can be classified on three types of techniques, including, symmetric key-ciphers, keyed hash function, and pseudo-random sequences. In order to secure data at the MAC layer, AES block cipher algorithm was proposed in the IEEE 802.15.4 \cite{95}. Yang et al. \cite{3} used keyed-hashing for message authentication code (HMAC) \cite{87} in order to achieve the integrity of the message and source data authentication. To compute HMAC, the scheme \cite{3} chooses a secret key for users  $K_j(\left|K_j\right|=128)$ for $U_j\in {\rm U}$ and $K_A(\left|K_A\right|=128)$ for $U_A$, then it sends $(s,K_j)$ and  $(d,K_A)$ to $U_j$ and $U_A$, respectively, where $s,d\in {{\rm Z}}_p$.

\subsection{Public-key primitives}
Public-key primitives can be classified on two types of techniques, including, public-key encryptions and signatures. We note that the public-key primitives is used mostly by privacy-preserving schemes.

Public-key encryption is potentially a good security solution when the number of nodes is very high. The schemes \cite{6,20,27} use the identity-based encryption \cite{38}. When an OBU with identifier ${ID}_i$ registers itself to the system, the proposed scheme uses the secret key $s$ to encrypt the real identifier ${ID}_i$ into a pseudo-ID ${PID}_i\ =\ {Encs}_s\ ({ID}_i{\rm ||\ }r_i)$, where $r_i$ is randomly chosen from ${{\rm Z}}^*_q$. Then, the OBU encrypts the message $M$ based on pseudo-ID  ${PID}_i$ , the current timestamp $T$, and the ephemeral key. The scheme \cite{7} uses the public key encryption with keyword search \cite{89,90}, which is takes as input the public key $PK$ and keywords $(w_1,w_2,\cdots ,w_l)$ that is associated with one document in order to output a ciphertext $C$. Then, it uses the trapdoor information $T_{{w'}_1,{w'}_2,\cdots ,{w'}_l}$ and  $Test(C,T_{{w'}_1,{w'}_2,\cdots ,{w'}_l})$.

Based on user attributes, Liang et al. \cite{12} adopted the attribute-based encryption \cite{93}, which outputs a ciphertext associated with the attribute set. Following a different approach and focusing on achieving efficient fine-grained filtering, Zhang et al. \cite{13} used the hidden vector encryption \cite{94,97}. Specifically, the filter creator in the scheme \cite{13} generates his fine-grained keyword filter as a vector $w=\left(w_1,\cdots ,w_l\right)\in {\{1,\cdots ,n\}}^l$. 

Dong et al. \cite{26} proposed the certificate less public key (CL-PKC) \cite{43,44} in order to achieve lightweight public key certificate management. The CL-PKC is specified by the following seven randomized algorithms: \textit{Setup, Partial-Private-Key-Extract, Set-Secret-Value, Set-Private-Key, Set-Public-Key, Encrypt, }and\textit{ Decrypt}. The \textit{Setup }algorithm is run by the key generating center (KGC), which outputs the system parameters $params$ and $master-key$. The \textit{Partial-Private-Key-Extract }algorithm returns a partial private key $D_A$using $params$, $master-key$, and an identifier for entity $A$,${ID}_A\in {\{0,1\}}^*$. The \textit{Set-Secret-Value }algorithm outputs $A$'s secret value $x_A$ using $params$ and ${ID}_A$. The \textit{Set-Private-Key }algorithm outputs the (full) private key $S_A$ using ${ID}_A$, and $x_A$. The \textit{Set-Public-Key }algorithm constructs the public key $P_A$ for entity $A$ using $params$ and $x_A$. The \textit{Encrypt }algorithm outputs a ciphertext $C$ using $P_A$ and ${ID}_A$. The \textit{Decrypt }algorithm returns a message $M$ using $params$, $C$, and $S_A$.
\begin{table}[h]
\centering
\caption{Game theoretic approaches used in privacy-preserving schemes}
\begin{tabular}{|p{1in}||p{1.2in}|p{1.7in}|} \hline 
\textbf{Scheme} & \textbf{Approach} & \textbf{Main results} \\ \hline 
ECPDR scheme \cite{4}\newline (2013) & Static game & Protect location privacy \\ \hline 
Scheme \cite{8}\newline (2012) & Cooperation game & Determine the optimal data forwarding strategy \\ \hline 
SSH scheme \cite{10}\newline (2011) & Sequence games & Prove that $IBE$ is IND-sPS-CPA secure under the k-SPDBDH assumption in the random oracle model \\ \hline 
PCS strategy \cite{19}\newline (2012) & Noncooperative information static game & Prove that the feasibility of the pseudonym changing at social spots (PCS) strategy \\ \hline 
SPECS scheme \cite{25}\newline (2011) & Static game & Protect message integrity and authentication \\ \hline 
FLIP protocol \cite{32}\newline (2010) & Sequence games & Demonstrate that the protocol is secure in the VANET scenarios \\ \hline 
The scheme \cite{35} \newline (2004) & Traceability game & Prove that the group signature scheme satisfies the requirements traceability \\ \hline 
\end{tabular}
\label{tab:Tab10}
\end{table}
Chim et al. \cite{21} proposed a proxy re-encryption \cite{39} that is based on the following algorithm: \textit{Setup}, \textit{KeyGen}, \textit{Encrypt}, \textit{Decrypt}, \textit{RKGen}, \textit{Reencrypt}. The \textit{Setup} algorithm is used in order to  output both the master public parameters and the master secret key. The \textit{KeyGen }algorithm outputs a decryption key ${sk}_{id}$ corresponding to identity $id$.  The \textit{Encrypt} algorithm  outputs $c_{id}$, the encryption of message under the specified identity. The \textit{Decryp t}algorithm  decrypts the ciphertext $c_{id}$ using the secret key ${sk}_{id}$. The \textit{RKGen }algorithm produces a re-encryption key ${rk}_{id_1\to {id}_2}$.  The \textit{Reencrypt} algorithm  outputs a ``re-encrypted'' ciphertext $c_{{id}_2}$.

Li et al. \cite{34} presented the non-interactive ID-based public-key cryptography \cite{48}, which is based on three following phases, namely: \textit{system setup}, \textit{user registration}, and \textit{authentication}, respectively. The \textit{system setup }phase outputs the public key $e$ in $Z^*_{\emptyset (N)}$ and a corresponding private key $d$ where $e*d\equiv 1(mod\emptyset \left(N\right))$. The \textit{user registration }phase outputs a secret key $s_i=e*{log}_g({ID}^2_i)(mod\emptyset \left(N\right))$ for the node $U_i$. The \textit{authentication }phase verifies $Y={({ID}^2_i)}^{r*s_j}={({ID}^2_i)}^{r*s_j}(mod\emptyset \left(N\right))$.

Homomorphic encryption \cite{85} is used by three schemes \cite{1,6} \cite{102}.  Rothblum in \cite{98} considers homomorphic encryption as a public-key scheme. Specifically, homomorphic encryption uses the public key $(N,g)$ and the corresponding private key $sk(\lambda ,\delta )$ in order to construct both the ciphertext $c$ and the plaintext $m$. The ciphertext $c=g^m\cdot r^nmod\ N$ where $r$ is a random number. The plaintext $m=L\left(c^{\lambda {\rm mod}N^2}\right)\cdot \delta \ {\rm mod}\ N$, where $(x)=(x-1)/N$ . The additive homomorphic property is as follows: $\left(m_1\right)\cdot E\left(m_2\right)=\left(g^{m_1}\cdot {r_1}^n\right)\left(g^{m_2}\cdot {r_2}^n\right)mod\ N^2=E(m_1+m_2)$.

To achieve anonymous authentication, the schemes in \cite{19,21,25,27,51} \cite{4,7} use the Boneh--Boyen short signature \cite{37}. In general, the following three algorithms specify a signature scheme: \textit{KeyGen}, \textit{Sign}, and \textit{Verify}. The \textit{KeyGen} algorithm outputs a random key pair $(PK,SK)$ . The \textit{Sign} algorithm constructs a signature $\sigma $ using a private key $SK$ and a message $M$. The \textit{Verify} algorithm verifies the signature and returns $valid$ or $invalid$. The Boneh--Boyen short signature \cite{37} is based on these three algorithms: \textit{Key Generation}, \textit{Signing}, and \textit{Verification}. The \textit{Key Generation }algorithm is same as \textit{KeyGen}. The \textit{Signing} algorithm outpust the signature  $(b,Sign(m))$ where $\sigma=H_1\left(SK,M\right)\in \{0,\ 1\}$ , $m=H_2(b,M) $ , $H_1$ and $H_2$ two hash functions. The \textit{Verification }algorithm outputs $valid$ if $Verify\left(PK,H_2\left(b,M\right),\sigma \right)=valid$. Note that there is the ID-based signature \cite{40} and the ID-based online/offline signature \cite{41} which are used by the scheme in \cite{22}.

Liang et al. in \cite{12} proposes a short group signature \cite{52} that is based on the following algorithm: \textit{Setup}, \textit{Join}, \textit{Sign}, \textit{Verify}, and \textit{Trace}, which could be executed by three parties: group manager, user, and verifier. The \textit{Setup }algorithm outputs the public parameters $PP$, the master key $MK$, and the tracing key $TK$, where  $PP=(g,h,Z)\in G\times G_q\times G_p$, $MK=z\in {{\rm Z}}^*_n$, $TK=q\in {\rm Z}$. $n=pq$ where $p,q$ are random primes and $G$ is a cyclic bilinear group and its subgroup $G_p$ and $G_q$ of respective order $p$ and $q$. $g\ $is a generator of $G$ and $h$ is a generator of $G_q$. The \textit{Join} algorithm construct the secret key $K_{id}=(s_{id},g^{\frac{1}{z+s_{id}}})$ , where $s_{id}$ is a user's identity $id$. The \textit{Sign }algorithm outputs signature $\sigma =({\sigma }_1,{\sigma }_2,{\sigma }_3,{\pi }_1,{\pi }_2)\in G^5$. The \textit{Verify} algorithm verifies $T_1=?e(h,{\pi }_1)$ and $T_2=?e(h,{\pi }_2)$.  The $Trace\ $algorithm trace the identity of signer using $PP$, $TK$, and $\sigma $. Moreover, note that  group signatures with verifier-local revocation \cite{35} are used in both the scheme in \cite{18} and the scheme in \cite{51}. In addition, a blind signature \cite{49,62} is used in the scheme \cite{34}, a schnorr signature algorithm \cite{45} is used in the scheme \cite{28}, and a ring signature \cite{42} is used in the scheme \cite{23}.

\subsection{Unkeyed primitives}

Unkeyed primitives can be classified on two types of techniques, including, hash functions and random sequences. The secure cryptographic hash functions \cite{36} are used in most privacy-preserving schemes for MSNs and VSNs, where the cryptographic hash function is used in order to check the integrity of a message. For example, modifying a message when transmitting can be proved by comparing the message hash value before and after transmission. Specifically, using a security parameter $\lambda $, a hashing function ${\mathcal H}:{\{0,1\}}^*\to {\{0,1\}}^{\lambda }$ is cryptographically secure if it satisfies three security properties, namely, pre-image-resistance, second pre-image resistance, and collision-resistance.

\subsection{Game theoretic approaches}

To prove the feasibility of privacy-preserving schemes in practice, researchers in the security field use various  mathematical tools such as  game theoretic approaches \cite{83}. Since the efficiency of cryptographic methods has already been proved, we note that only a small fraction of  privacy-preserving schemes that we have examined, used game theoretic approaches, which are summarized in Tab. \ref{tab:Tab10}. More precisely, Ferrag et al. in \cite{4} use the static game \cite{99} in the ECPDR scheme to protect location privacy. Liang et al. in \cite{8} uses the cooperation game to determine the optimal data forwarding strategy. The SSH scheme \cite{10} uses the sequence games to prove that $IBE$ is IND-sPS-CPA is secure under the k-SPDBDH assumption in the random oracle model. The pseudonym changing at social spots (PCS) strategy \cite{19} uses non-cooperative information static game \cite{99} to prove the feasibility of PCS. Similarly to the scheme \cite{4}, Chim et al. in \cite{25} use the static game to protect message integrity and authentication. The sequence games is used by the FLIP protocol \cite{32} to demonstrate that the protocol is secure in the VANET scenarios. Finally, the scheme \cite{35} use the traceability game to prove that the group signature scheme satisfies the requirements traceability. For more details, we refer the reader to the survey \cite{83}.

\begin{center}
\topcaption{Summary of privacy-preserving schemes for MSNs (Published between 2011 and 2016)}
\end{center}
{\small \begin{supertabular}{|p{0.4in}||p{0.8in}|p{1in}|p{1in}|p{1.4in}|p{1.4in}|} \hline 
\textbf{Scheme} & \textbf{Network model} & \textbf{Privacy model} & \textbf{Goals} & \textbf{Main phases} & \textbf{Performances (+) and limitations (-)} \\ \hline 
Liang et al. \cite{1}\newline \newline (2013) & Each user has a profile with dimension vector to find the targeting user & Non-anonymity;\newline Conditional anonymity;\newline Full Anonymity & Users compare their profiles while not disclosing the profiles & Explicit comparison-based approach;\newline Implicit comparison-based approach;\newline Implicit predicate-based approach & + Anonymity break period\newline + Anonymity risk level\newline -Limited analysison the impact of ``='' on the anonymity \\ \hline 
Yang et al. \cite{3}\newline (2016) & Specified by a trusted key distribution center with a semi-trusted data processing center & Privacy preservation the profiles of users  & Minimize the personal profiles disclosure & Cosine similarity matching;\newline Weighted $l_1$-norm matching & + Computation complexity\newline + Average running time vs. number of profile items\newline -The layer routing is not considered \\ \hline 
Ferrag et al. \cite{4}\newline (2013) & Each user has a sociality strength in a set of social hotspots & Immutability;\newline Transparency;\newline Accountability; & Improve routing by privacy preservation & Node certificate updating;\newline Message signature and verification;\newline Response Requested;\newline Demand Response & + Black hole detection rate\newline + Transmission delay\newline -Limited analysis with few adversaries \\ \hline 
Liang et al. \cite{5}\newline (2014) & Multiple vendors offering similar services to users where each vendor is equipped with a wireless communication device & Trust evaluation & Detect and prevent the sybil attacks & Structured reviews;\newline Synchronization tokens;\newline Review generation and submission & + Provides a good security analysis of sybil attacks\newline - No comparison with other systems  \\ \hline 
Ferrag et al. \cite{6}\newline (2014) & Peer-to-peer node community with a large number of mobile users & Transparency;\newline Impersonator resistance & Provides the strong privacy-preservation of message;\newline Provides the evolution of users' certificates & Detecting attacks;\newline Response Requested;\newline Demand Response;\newline Certificate evolution & + Detectreq reporting delay\newline + Transmission delay\newline - Need of large Detectreq reporting \\ \hline 
Ferrag et al. \cite{7}\newline (2016) & Each user has a sociality strength & Immutability;\newline Transparency;\newline Accountability & Detect and prevent the wormhole attacks & Peer registration phase;\newline Document forwarding phase;\newline Detection, verification and avoidance & + Hole link detection accuracy\newline + Transmission delay\newline -Many assumptions needed to understand implementation \\ \hline
Liang et al. \cite{8}\newline (2012) & Each user has a sociality strength and a morality state in a set of social hotspots & Location privacy;\newline Identity privacy & Privacy preservation and cooperative data forwarding & Privacy-preserving route-based authentication;\newline Proximity measurement;\newline Morality-driven data forwarding & + Sociality strength\newline + Cooperation effect\newline + Delivery ratio in games\newline - Diverse behavior models is not considered \\ \hline
Liang et al. \cite{9}\newline (2012) & Patients and doctors communicate through health social networks;\newline Each user has a social-active factor & Privacy of information shared & Attribute-oriented authentication;\newline Attribute-oriented transmission & Attribute initialization;\newline Attribute-oriented authentication;\newline Attribute-oriented transmission & + Impact of social-active factor\newline + Attribute-oriented transmission\newline -No comparison with other methods \\ \hline 
Lu et al. \cite{10}\newline (2011) & Patients and doctors communicate through mobile health social networks & Identity privacy;\newline Impersonator resistance & Provides a secure same-symptom-based handshake & System setup algorithm;\newline Patient joining algorithm;\newline Patients same-symptom-based handshaking algorithm & + Average delivery ratio\newline + Average reporting delay\newline -No comparison with other methods \\ \hline 
Hameed et al. \cite{11}\newline (2011) & Each user has a sociality with his e-mail & Trust evaluation & Prevention of Spam transmission & Community formation;\newline Trust management;\newline Gate Keeper selection;\newline Spam report handler & + Performance of email filtration\newline -Limited analysis, no comparison with prevent spam transmission schemes available. \\ \hline 
Liang et al. \cite{12}\newline (2011) & Patients and doctors communicate through health social networks & Identity privacy;\newline Privacy of information shared & Enhancing availability; \newline Ensuring unlinkability of the transactions & Registration;\newline Emergency call generation;\newline Emergency call verification & + Decryption efficiency\newline + Revocation efficiency\newline -No consideration the decentralized emergency response system \\ \hline 
Zhang et al. \cite{13}\newline (2015) & Each mobile user has a sociality in ad hoc network with local stores. & Identity privacy\newline  & Develop a personalized fine-grained filtering & Social-assisted filter distribution;\newline Coarse-grained and fine-grained filters;\newline Merkle Hash tree-based filter authentication and update & + Can efficiently update the distributed filters\newline -Many assumptions needed to understand implementation \\ \hline 
Li et al. \cite{50} \newline (2016) & Each mobile user has a sociality with the sensitive data or demographics of the target. & Privacy of location sharing & Provide different privacy controls & Maximum common trace based inference approach;\newline Machine learning based inference approach & + Comparison of shared mobility and ground truth traces\newline - No comparison with other frameworks \\ \hline 
Luo et al. \cite{102}\newline (2016) & Each user has a personal profile in the course of friendship discovery & User privacy preservation & Ensures high level privacy of user profile information & One-hop friend discovery matching;\newline Multi-hop friend discovery matching & + Communication overhead\newline + Comparison with other methods\newline - No consideration the identity privacy and location privacy \\ \hline 
\end{supertabular}}

\section{Privacy-preserving schemes for MSNs}\label{sec:privacy-preserving-schemes-for-msns}
In this section, we in-detail examine fourteen privacy-preserving schemes developed for or applied in the context of MSNs. Based on the network model, we classify these schemes in six categories, including, social profile, social morality, social routing, social health, location-based services, and service-oriented sociality. In addition, these schemes as shown in Tab. XI are published between 2011 and 2016.

\subsection{Network model with social profile}

In \cite{1}, Liang et al. considers that each user has a profile represented by a distinct dimension vector that can be used to find the targeting user. Specifically, the work in \cite{1} presents a scheme, called PPM, which can preserve three privacy models, including, 1) non-anonymity, 2) conditional anonymity, and 3) full anonymity. With the PPM scheme, users can compare their profiles while not disclosing the profiles. The PPA uses three main phases, namely, explicit comparison, implicit comparison, and implicit predicate. The PPM scheme is efficient in terms of anonymity break period and anonymity risk level, but the article fails to provide a detailed analysis on the impact of ''='' on the anonymity. Privacy preserving of the profiles of users is an important topic as identified in \cite{3}. Yang et al. in \cite{3} characterized a MSN by a trusted key distribution center with a semi-trusted data processing center and he proposed a privacy preserving protocol called SFPM. In order to minimize the personal profiles disclosure, the SFPM protocol uses two main phases of matching, including, 1) cosine similarity and 2) weighted $l_1$-norm. The SFPM protocol is efficient in terms of computation complexity and average running time vs. number of profile items, but the layer routing is not considered. In another recent work \cite{102} Luo et al. proposes a privacy-preserving multi-hop profile-matching protocol for proximity-based mobile social networks.

\subsection{Network model with social morality}

In \cite{8}, authors state that each user has a sociality strength and a morality state in a set of social hotspots. Specifically, Liang et al. \cite{8} developed a protocol for privacy preservation and cooperative data forwarding, which can protect both the location privacy and the identity privacy of the user. This protocol uses three main phases, including, 1) privacy-preserving route-based authentication, 2) proximity measurement, and 3) morality-driven data forwarding. In addition, this protocol is efficient in terms of sociality strength, cooperation effect, and delivery ratio in games, but diverse behavior models are not considered.

\subsection{Network model with social routing}

The routing protocol in social ad hoc networks is a principal element to efficiently route the produced social data. The works in \cite{4,6,7} consider the sociality in routing protocols as the OLSR protocol \cite{70} and the AODV protocol \cite{71}. Ferrag et al. in \cite{4} developed a scheme, called ECPDR, in order to improve routing by incorporating the privacy preservation dimension. The ECPDR scheme can provide immutability, transparency, and accountability. For detecting attacks, the ECPDR scheme uses four main phases, including,1) node certificate updating, 2) message signature and verification, 3) response requested, and 4) demand response. In addition, the ECPDR scheme is efficient in terms of black hole detection rate and transmission delay, but gives a limited analysis with few adversaries. Ferrag  et al. \cite{6} considering a peer-to-peer node community with a large number of mobile users proposes a novel scheme, called SDPP. Based on the certificate evolution phase, the SDPP scheme can provide transparency and impersonator resistance. The SDPP scheme is efficient in terms of reporting delay and transmission delay, but leads to the creation of large reports. Ferrag et al. \cite{7} focuses on detecting and preventing the wormhole attacks and proposes a scheme called EPSA. The EPSA scheme is based on three main phases, including, 1) peer registration, 2) document forwarding, and 3) detection, verification and avoidance. The ESPA scheme is efficient in terms of hole link detection accuracy and transmission delay, but makes too many assumptions regarding the network characteristics.

\subsection{Network model with social health}

The health social networks (HSNs) which are essential for the communication between patients and doctors demand highly efficient privacy-preserving schemes. In \cite{9}, Liang et al. considers that each user has a social-active factor in HSN. The HealthShare scheme in \cite{9} is proposed in order to provides privacy of information shared, which is devided in three main phases including, 1) attribute initialization, 2) attribute-oriented authentication, and 3) attribute-oriented transmission. The HealthShare scheme is efficient in terms of the impact of social-active factor and the attribute-oriented transmission, but the article doesn't present a  comparison of the proposed mechainsm with other methods. In \cite{12}, Liang et al. developed a scheme, called PEC, for HSN. The PEC scheme ensures unlinkability of the transactions and it is  based on two main algorithms; 1) emergency call generation and 2) emergency call verification.The PEC scheme is efficient in decryption and revocation, but needs the consideration of the decentralized emergency response system. Similarly to \cite{9} and \cite{12}, in \cite{10}, Lu et al. proposed a scheme, called SSH, which targets  mobile users in HSNs. The SSH scheme provides a secure same-symptom-based handshake based on two main phases, including,1) patient joining and 2) patients same-symptom-based handshaking. In addition, the SSH scheme is efficient in delivery ratio and reporting delay, but authors dont present a thorough comparison of their system with othersimilar methods.

\subsection{Network model with location-based services}

The geosocial networking is a new concept in the topic of social networking, where each mobile user has a sociality whcih is associated with some sensitive data or the demographics of the target. In \cite{50}, an interesting recent work considers the demographics (e.g., age, gender, education) in MSN, and based on these the authors proposed a new set of attacks that can infer the demographics. Specifically, in order to provide full privacy of location sharing, one needs to combine the following approaches in \cite{50}: 1) maximum common trace based inference approach and 2) machine learning based inference approach. The work in \cite{50} presents a good comparison of shared mobility and ground truth traces. In addition, the work in \cite{50} developed a framework, called SmartMask, for the protection of location privacy, which needs to be compared with other frameworks in the future in order to test its efficiency. The idea of social spot in MSN proposed in the VSLP protocol \cite{72} can be applied in this category.

\subsection{Network model with service-oriented sociality}

Liang et al. \cite{5} developed a system, called TSE, which considers  multiple vendors offering similar services to users where each vendor is equipped with a wireless communication device. The TSE system provides a trust evaluation mechanism and can detect and prevent a sybil attacks. The TSE system is based on three main phases, namely, structured reviews, synchronization tokens, and review generation and submission. The work in \cite{5} provides a good security analysis of the mechanism against sybil attacks, but lacks comparison with other systems. Similarly to the TSE scheme \cite{5}, the LENS scheme \cite{11} also provides a trust evaluation and prevention mechanism  against spam transmission. Zhang et al. \cite{13} developed a personalized fine-grained filtering scheme, called PIF. The PIF scheme considers that each mobile user has a sociality in an ad hoc network with local stores. The PIF scheme is based on three main phases, namely, social-assisted filter distribution, coarse-grained and fine-grained filters, and merkle hash tree-based filter authentication and update. In addition, the PIF scheme can efficiently update the distributed filters  but makes too many assumptions regarding the network characteristics. 

\newpage
\begin{center}
\topcaption{Summary of privacy-preserving schemes for VSNs (Published between 2008 and 2016)}
\end{center}
{\small \begin{supertabular}{|p{0.4in}||p{0.8in}|p{1in}|p{1in}|p{1.4in}|p{1.4in}|} \hline
\textbf{Scheme} & \textbf{Network model} & \textbf{Privacy model} & \textbf{Goal} & \textbf{Main phases} & \textbf{Performances (+) and limitations (-)} \\ \hline 
Lu et al. \cite{18}\newline \newline (2012)\newline \newline  & A typical location based services in VANET & Forward secrecy~;\newline Backward secrecy~;\newline Collusion resistance\newline  & Support a privacy-preserving authentication in the vehicle-user-joining phase~;\newline Enable vehicle users to autonomously update the session key & Location based services settings~;\newline Vehicle user joining~;\newline Vehicle user departure & \textbf{+}Key update delay\newline \textbf{+}Key update ratio\newline -Compared only with the traditional key update \\ \hline 
Lu et al. \cite{19}\newline (2012) & VANET with a collection of social spots & Location privacy\newline  & Facilitate vehicles to achieve high-level location & Key generation~;\newline Pseudonym self-delegated generation~;\newline Conditional tracking & \textbf{+}Anonymity set size\newline \textbf{+}Location privacy gain\newline + Feasibility is proved using game-theoretic techniques\newline - Limited analysis with the threat model \\ \hline 
Lu et al. \cite{20}\newline (2010) & VANET with a large number of parking spaces & Conditional privacy preservation & Develop an intelligent parking for large parking lots~;\newline Support a privacy-preserving of the drivers & Real-time parking navigation~;\newline Intelligent antitheft protection~;\newline Friendly parking information dissemination & \textbf{+}Searching time delay\newline - No comparison with other scheme in term of coverage ratio \\ \hline 
Chim et al. \cite{21}\newline \newline (2014) & VANET  & Identity privacy-preserving\newline Traceability & Guide vehicles to desired destinations in a distributed manner~;\newline Support a privacy-preserving of the drivers & Generation of anonymous credentials~;\newline Activation and requesting for master key~;\newline Requesting for anonymous credential~;\newline Requesting for navigation service~;\newline Navigation request and reply propagation~;\newline Verification of RSUs' hop information~;\newline Guiding to destination & \textbf{+}Processing delay\newline \textbf{+}Reduction in travelling time\newline + Analysis on time complexity\newline - Limited analysis with the threat model \\ \hline 
Huang et al. \cite{22}\newline (2012) & VANET & Identity privacy-preserving~;\newline Traceability & Solving the issues of authentication and privacy in VANET & V2R and R2V authentication~;\newline V2V authentication~;\newline Cross-RSU V2V authentication & \textbf{+}Storage requirement and computation\newline - No threat model presented \\ \hline 
Xiong et al. \cite{23}\newline (2012) & VANET with a member manager & Multi-level anonymity & Support a multi-level conditional privacy preserving & OBU safety message generation~;\newline Message verification~;\newline OBU fast tracing & \textbf{+}Storage requirements\newline \textbf{+}Computational overheads\newline - Limited analysis with the threat model \\ \hline 
Ying et al. \cite{24}\newline (2013) & VANET & Anonymity  & Support a privacy preserving of broadcast message & Vehicle registration~;\newline Hash chain generation~;\newline Data transmission & \textbf{+}Average link layer delay\newline \textbf{+}Average data packet delay\newline - Location privacy is not considered \\ \hline 
Chim et al. \cite{25}\newline \newline (2011) & VANET & Identity privacy preserving~;\newline Traceability~;\newline Revocability & Satisfy the privacy requirement based only on two shared secrets & Initial handshaking~;\newline Message signing~;\newline Batch verification~;\newline Real identity tracking and revocation~;\newline Group key generation~;\newline Group message signing and verification & \textbf{+}Data transmission\newline \textbf{+}Invalid batch successful rate\newline - Location privacy is not considered\newline  \\ \hline 
Dong et al. \cite{26}\newline (2011) & VANET with a lite certificate authority & Location privacy~;\newline Anonymity & Solving the issues of authentication and privacy in VANET & Lite-CA-based public key cryptosystem~;\newline Identity-based public key cryptosystem & \textbf{+}Encryption cost comparison\newline \textbf{+}Computational cost comparison\newline - Limited consideration of routing protocols. \\ \hline 
Huang et al. \cite{27}\newline \newline (2011) & VANET & Conditional privacy preservation~;\newline Anonymity & Solving the issues of authentication and privacy in VANET such as low pseudonym generation latency, high scalability, and easy revocation & Registration and generation~;\newline Extraction~;\newline Encryption and decryption~;\newline Revocation & \textbf{+}Protocol latency analysis\newline \textbf{+}Comparison of search times for revocation\newline - Mobility models are not considered \\ \hline 
Sun et al. \cite{28}\newline \newline (2010) & VANET & Backward privacy~;\newline Conditional anonymity~;\newline Nonrepudiation~;\newline Identity revocation & Solving the issues of authentication and privacy in VANET & RSU certificate issuing~;\newline Vehicle pseudonymous certificate issuing~;\newline Vehicle pseudonymous certificate updating~;\newline Identity revocation~;\newline Message signature and verification. & \textbf{+}Revocation overhead\newline \textbf{+}Certificate updating overhead\newline \textbf{+}Authentication overhead\newline + Efficient compared to other schemes\newline - Location privacy is not considered \\ \hline 
Lin et al. \cite{29}\newline (2011) & VANET with a collection of social spots & Location privacy~;\newline Vehicle conditional privacy preservation & Achieving receiver-location privacy preservation in VANETs & Packet sending~;\newline Social-tier dissemination~;\newline Packet receiving & \textbf{+}Average delivery ratio\newline \textbf{+}Packet average delay\newline -No comparison with other protocols  \\ \hline 
Lu et al. \cite{30}\newline \newline (2011)\newline  & VANET with a collection of social spots (including small social spot and large social spot) & Location privacy~;\newline Anonymity & Achieving the location privacy based on pseudonyms changing technique & Pseudonym changing at small social spot~;\newline Pseudonym changing at large social spot & \textbf{+}Anonymity set size\newline -No comparison with other methods \\ \hline 
Lu et al. \cite{31}\newline (2010) & VANET with a collection of social spots & Receiver-location privacy\newline  & Achieving the location privacy based ``Sacrificing the Plum Tree for the Peach Tree'' tactic  & Packet generation~;\newline Packet forwarding~;\newline Packet receiving & \textbf{+}Average packet delivery ratio\newline \textbf{+}Average packet delay\newline -No comparison with other protocols \\ \hline 
Lu et al. \cite{32}\newline (2010) & VANET with not include RSUs & Identity privacy~;\newline Location privacy~;\newline Interest privacy & Facilitate vehicles to communicate the common interest~;\newline Protects the interest privacy from other vehicles who don't have the same interest & Privacy preserving finding like-minded vehicle on the road & \textbf{+}Average delay for finding the like-minded vehicle\newline \newline - Limited analysis with the threat model \\ \hline 
Lu et al. \cite{33}\newline (2010) & VANET with a social degree of an intersection vertex & Conditional privacy preservation & Optimizing vehicular DTN with RSU assistance~;\newline Resisting privacy-related attacks on vehicle DTN nodes~;\newline Achieving conditional privacy preservation & Opportunistic RSU-aided packet forwarding & \textbf{+}Average delivery ratio\newline \textbf{+}Packet average delay\newline \newline - Limited consideration of routing requirements \\ \hline 
Li et al. \cite{34}\newline \newline (2008) & VANET & User privacy preservation & Achieving conditional privacy preservation based on a lightweight authenticated key establishment scheme & Handling new vehicles, roadside devices, and service providers~;\newline Scenario 1: secure communications between vehicles~;\newline Scenario 2: secure communications between vehicles and roadside devices~;\newline Scenario 3: a secure and efficient communication scheme with privacy preservation & \textbf{+}Computational overhead\newline \textbf{+}Communication overhead\newline \textbf{+}Storage overhead\newline \newline - Limited consideration of routing requirements \\ \hline 
Rabieh et al. \cite{118} & VANET with a centralized authority &  Interest privacy & Protects the interest privacy from other vehicles who don't have the same interest & Chatting request packet~;\newline Chatting response packet~;\newline Degree of interest verification~;\newline Interest revocation & \textbf{+}Computational overhead\newline \textbf{+}Communication overhead\newline - Limited analysis with the threat model\newline -No comparison with other protocols\newline - Mobility models are not considered\newline - Location privacy is not considered\textbf{} \\ \hline 
Yu et al. \cite{51}\newline \newline (2016) & VANET with data center and a collection of social spots (including Global Social Spot and Individual Social Spot ) & Location privacy & Exploit the meeting opportunities for pseudonym changing~;\newline Improve the location privacy preservation & System initialization and key generation~;\newline Group join~;\newline Pseudonyms exchanging~;\newline RSU signing protocol~;\newline Group leaving~;\newline Revocation protocol~;\newline Conditional tracking & \textbf{+}Global pseudonym entropy of the entire VSN\newline \textbf{+}Expected and actual pseudonym entropy of a target vehicle\newline + Comparison with existing schemes\newline + Analysis with the threat model\newline -Many assumptions needed to understand implementation\newline  \\ \hline 
\end{supertabular}
}
\begin{table}[h]
\centering
\caption{Models of social spots}
\begin{tabular}{|p{1.2in}||p{1.1in}|p{1.4in}|} \hline 
\textbf{Scheme} & \textbf{Model} & \textbf{Idea} \\ \hline 
SPF protocol \cite{31} (2010)\newline STAP protocol \cite{29} (2011) & Social spot integrated with RSU & Store packets in packet forwarding \\ \hline 
PCS strategy \cite{19} (2012)\newline Scheme \cite{30} (2011) & Small social spot and large social spot & Vehicle changes its pseudonym at social spot  \\ \hline 
MixGroup \cite{51} (2016) & Global social spot and individual social spot & Exploit the meeting opportunities for pseudonym changing \\ \hline 
\end{tabular}
\label{tab:Tab13}
\end{table}
\section{Privacy-preserving schemes for VSNs}\label{sec:privacy-preserving-schemes-for-vsns}

In this section, we in-detail examine nineteen privacy-preserving schemes developed for or applied in the context of VSNs. Based on the network model, we classify these schemes in three categories, including, social spots, location-based services, and service-oriented sociality. In addition, these schemes as shown in Tab. XII are published between 2008 and 2016.

\subsection{Network model with social spots}

The idea of placing social spots in VANET networks has emerged as an important research area, which is referred to as the locations where many vehicles will visit, for example, a sports complex, or a parking \cite{74}. As shown in Tab. \ref{tab:Tab13}, there are three models of social spots, including, 1) social spot integrated with RSU, 2) small social spot and large social spot, and 3) global social spot and individual social spot. In \cite{19}, the work developed a strategy called PCS, which consider the VANET with a collection of social spots in order to facilitate vehicles to achieve high-level location preservation. To preserving the location privacy, the PCS strategy uses three main phases, namely, key generation, pseudonym self-delegated generation, and conditional tracking. The PCS strategy is efficient in terms of anonymity set size and location privacy gain. In addition, the feasibility is proved using game-theoretic techniques but the authors conducted a  limited analysis of the different threat models.

Similarly to the PCS strategy, Lin et al. in \cite{29} developed a protocol called STAP. The STAP protocol considers social-tier-assisted VANET network. With the assistance of social spot, STAP is not only very efficient in terms of packet delivery ratio and packet average delay, but also can preserves the location privacy. In order to unveil the asymptotic performance limits in MSN, the work in \cite{73} can be applied in this category. In \cite{30}, Lu et al. proposed a social spot based pseudonyms changing technique, which considers VANET with a collection of social spots (including small social spot and large social spot). In addition, Lu et al.  \cite{30} proposed a protocol called SPF. The SPF protocol achieving the location privacy based ``Sacrificing the Plum Tree for the Peach Tree'' tactic. The SPF protocol is efficient in terms of average packet delivery ratio and average packet delay, but lacks comparison with similar protocols.

The MixGroup scheme in \cite{51} is a recent interesting work, which considers a VANET with data center and a collection of social spots (including Global Social Spot and Individual Social Spot). With the assistance of social spot, MixGroup is not only very efficient for pseudonym changing, but also can improve the location privacy preservation of the users.

\subsection{Network model with location-based services}

The location-based services play an important role in ad hoc social networks \cite{75,76}. In \cite{18}, Lu et al. developed a scheme called DIKE, which considers a typical location-based service in VANET. Using the cooperative key update with V-2-V communication, the DIKE scheme can preserve forward secrecy, backward secrecy, and collusion resistance. During each key update procedure, DIKE is efficient in terms of key update delay and key update ratio. 

The applications services such as parking are very important in VANET \cite{77,78}. Lu et al. in \cite{20} consider VANET with a large number of parking spaces. The work \cite{20} developed an intelligent parking method that can be applied to large parking lots, which can support  privacy-preserving of the drivers information. In addition, the work \cite{20} is efficient in term of searching time delay. In a similar work Xiong et al. in \cite{23} considering a  VANET with a member manager,  developed a protocol to support  multi-level conditional privacy. This protocol is efficient in terms of storage requirements and computational overhead..

For solving the issues of authentication with privacy, Dong et al. in \cite{26} proposed a scheme, called EP2DF, which considers VANET with a lite certificate authority. EP2DF is based on two cryptosystems, including, 1) lite-CA-based public key cryptosystem and 2) identity-based public key cryptosystem. The lite-CA-based public key cryptosystem is proposed specially to achieve lightweight public key certificate management. In addition, EP2DF is efficient in terms of encryption cost comparison and computational cost comparison. Related to EP2DF, Sun et al. \cite{28} proposed a scheme called PASS, which it is efficient in terms of revocation overhead, certificate updating overhead, and authentication overhead. In addition, PASS is efficient compared to other schemes like ECPP scheme \cite{47} and DCS scheme \cite{81}.

\subsection{Network model with service-oriented sociality}

The modeling of service-oriented sociality is based on a set of intersection nodes. Lu et al. in \cite{33} proposed a scheme called SPRING and introduces the Social Degree of an intersection vertex in VANET, which is used for the optimal deployment of RSUs. In addition, the work in \cite{33}  proposes a method that optimizes vehicular DTN with RSU assistance. The SPRING scheme is efficient in terms of average delivery ratio and packet average delay. On the other hand, the FLIP scheme in \cite{32} considers a VANET without the assistance of RSUs, which not only facilitates vehicles to communicate any common interest but can also protect the interest privacy from other vehicles who don't share the same interest. FLIP is efficient in terms of average delay for finding the like-minded vehicle.

Since the GPS integration into vehicles, each vehicle in a VANET can find the geographically shortest route based on a local map database \cite{79}. Using social interplay, the NextCell scheme \cite{80} can predict the location of a user from cell phone traces. However, obtaining  an accurate position with privacy-preserving is very important for vehicle applications. Chim et al. in \cite{21} proposed VSPN scheme, which can guide vehicles to desired destinations in a distributed manner and support  privacy-preserving of the drivers. VSPN is efficient in terms of processing delay and reduction in travelling time. 

Based on the VANET communications, Li et al. in \cite{34} develop a schema called SECSPP and characterize the security in VANET by three scenarios, including, 1) secure communications between vehicles, 2) secure communications between vehicles and roadside devices, 3) a secure and efficient communication scheme with privacy preservation. SECSPP scheme is efficient in terms of computational overhead, communication overhead, and storage overhead. The scheme in \cite{22} using a similar idea, proposed three authentication scenarios, namely, V2R and R2V authentication, V2V authentication, and cross-RSU V2V authentication.

Based on the state transition diagram in \cite{27} for pseudonym generation, PACP scheme can solve the issues of authentication and privacy in VANET such as low pseudonym generation latency, high scalability, and easy revocation. In addition, PACP is efficient in terms of protocol latency analysis and comparison of search times for revocation. Related to PACP, SPECS scheme in \cite{25} is efficient in terms of data transmission and invalid batch successful rate. On the other hand, the PPBMA scheme in \cite{24} considers the link layer in privacy preserving broadcast message. PPBMA is efficient in terms of link layer delay and data packet delay.

\begin{figure}[h]
\centering
\includegraphics[width=1\linewidth]{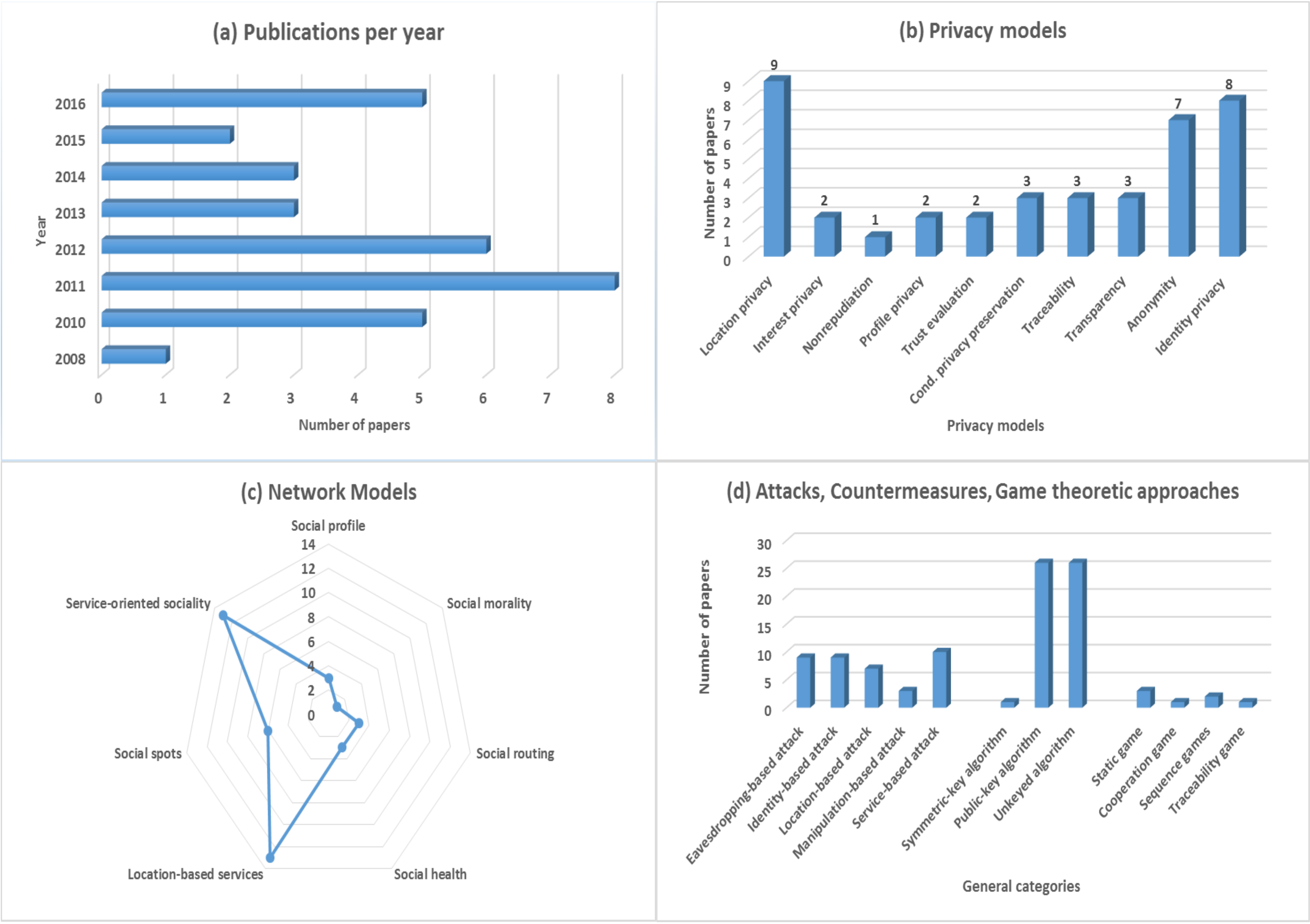}
\caption{(a) Publication per year, (b) Number of papers vs. privacy models, (c) Number of papers vs. network models, and (d) Number of papers vs. general categories}
\label{fig:Fig7}
\end{figure}

\section{Recommendations for further research}\label{sec:recommendations-for-further-research}
The average date of publication of the surveyed papers is 2012, as shown in Fig. \ref{fig:Fig7} (a). However, 70\% of these papers focus on three privacy models, namely, location privacy, identity privacy, and anonymity, as shown in Fig. \ref{fig:Fig7} (b). For network models in MSNs and VSNs, most papers use service-oriented social, social spots, and location-based services, as shown in Fig. \ref{fig:Fig7} (c). As shown in Fig. \ref{fig:Fig7} (d), manipulation-based attacks are less considered and 95\% of the surveyed papers use both public-key algorithms and unkeyed algorithms. In addition, only seven papers use four types of game theory approach, namely, static game, cooperation game, sequence games, and traceability game. 

In the remainder of this section, we will discuss four open issues for ad hoc social networks, including, privacy preserving methods, interdependent privacy, combination of privacy metrics, and identification of areas of vulnerability.

\subsection{Privacy-preserving methods}

We recommend three privacy-preserving methods, namely, 1) privacy-preserving energy consumption, 2) privacy preservation for V2G social networks, and 3) privacy preservation for social internet of vehicles. 

\begin{itemize}
\item  \textbf{Privacy-preserving energy consumption: }Privacy-preserving energy consumption is an open issue in ad hoc social networks. There are several research works addressing energy problems in ad hoc networks \cite{103}. Based on optimal numbers of clusters, Ali et al. in \cite{103} proposed an algorithm called MOPSO to manage the resources in order to make the MANET energy-efficient. A recent idea of Maglaras et al. in \cite{104} can improve the use of social clusters based on semi-markov processes. Therefore, how to manage energy consumption under social clustering of nodes in ad hoc networks? Hence, privacy-preserving energy consumption is one of the future works.

\item  \textbf{Privacy preservation for V2G social networks:} A recent survey published in 2016 \cite{105} review the state of the art of privacy-preserving schemes for V2G networks in smart grid, none of them carries study for the social characteristics in V2G networks. The future works addressing the limitations of privacy-preserving schemes for V2G networks will have an important contribution for V2G social networks.

\item  \textbf{Privacy preservation for social internet of vehicles:} Privacy preservation for social internet of vehicles (SIoV) is an open issue that we are working on \cite{107}. Since the SIoV is a combination of social and vehicular networks, the future works addressing the limitations from both domains will have an important contribution for the SIoV.
\end{itemize}

\subsection{Interdepedent privacy}

Interdependent privacy refers to situations where the privacy of individual users is affected by the decisions of others \cite{108}. Especially in social networks, where the interaction among different entities is constant and when talking for VSNs these entities are unknown \cite{107} the interdependent privacy is playing a key role. One excellent example of privacy interdependence is the Facebook application platform. How well a user can protect his privacy from third party developers depends not only on his decisions, but also on the decisions of his friends. New applications that are based on the social concept of vehicular networking, like Navitweet \cite{109} and Caravan Track \cite{110} combined with traditional OSNs can create new privacy threats for the users. 

\subsection{Combination of privacy metrics}

New privacy-enhancing technologies and those resulting from combinations of existing technologies need to be evaluated thoroughly to make sure that they provide an adequate amount of privacy. Because the VSN is a combination of social and vehicular networks, evaluations need to use a selection of privacy metrics from both domains \cite{111}.

\subsection{Identification of areas of vulnerability}

Since we are moving to the ear of IoT, a data breach from a system can be initiated from an attack that has occurred on another system that is somehow interconnected with it. For example in a VSN, an attack to a weak node can initiate malware propagation to the rest of the network, which is very difficult to detect and stop \cite{112}. A smart vehicle with an out of date software \cite{113} or an unprotected RSU can play the role of the weak node when an attack can be initiated. The US Department of Defense use the CARVER assessment method to determine criticality and vulnerability in enemy infrastructures. In a MSN/VSN the identification of critical nodes could help apply different privacy technologies according to how vulnerable each entity is.  Therefore, research in vulnerability identification is needed in order to handle the complexity and heterogeneity of modern Ad Hoc social networks. 

Although new sophisticated privacy metrics are needed, along with thorough analysis of the system in order to spot the weak players, no privacy metric is efficient if the human factor is neglected \cite{114}. Organizations and society continue to be affected by both regular and similar cyber security breaches. These breaches pertain to technical implementations as well as routine processing of confidential electronic information. Despite this range of activities, it has been proven that half of these have human error at their core \cite{115} and human aspects of cyber security must be taken into account when building new security and privacy mechanisms.  Recently the idea of quantum cryptographic approaches for privacy preserving in cloud \cite{vidya2016quantum} and wireless systems \cite{nomula2016multi} were introduced. Quantum cryptographic techniques provide an additional layer of security and privacy preservation of the system. Therefore, research in quantum privacy approaches is needed to handle the various attacks that MSNs and VSNs face. Although there exist simulator framework that combine many different simulators in order to better represent reality \cite{kosmanos2016}, the evaluation of the proposed metrics on real environments is also an open issue especially for VSNs where real deployments are limited.

\section{Conclusion}\label{sec:conclusion}

In this article, we surveyed the state-of-the-art of privacy-preserving schemes for both MSNs and VSNs. We presented the major privacy models, including, location privacy, identity privacy, anonymity, traceability, interest privacy, backward privacy, and content oriented privacy. We also presented the major threats including, identity-based attacks, location-based attacks, eavesdropping-based attacks, manipulation-based attacks, and service-based attacks. We reviewed the countermeasures and game theoretic models proposed for MSN and VSN privacy preservation. We  presented a side-by-side comparison in a tabular form for the current state-of-the-art of privacy-preserving schemes (thirty-three) proposed for MSN and VSN.  Privacy preservation in MSNs and VSNs remains a challenging problem since adversaries can find different ways for exploiting vulnerabilites of the system. As we move to the IoT era, privacy preservation of a network cannot be treated in isolation but interedepedencies among users and networks must be taken into account. The correct identification of vulenrabilites of the system and the combincation of privacy metrics can improve the protection of the system, but no countermeasure can be effective if the human factor is neglected.

%\appendix\cite{Proof of the Zonklar Equations}
% or


\ifCLASSOPTIONcaptionsoff
  \newpage
\fi



\bibliographystyle{IEEEtran}
\bibliography{Bibsurvey}


%\vfill


\end{document}